\documentclass[aps,prd,twocolumn,showpacs,superscriptaddress,groupedaddress]{revtex4}  % for review and submission
\pdfoutput=1 
 \def\LineNumbers{false}	
\usepackage{ifthen}
\usepackage{lineno}
\usepackage{slashed}
\usepackage{epsfig}
\usepackage{amsmath}
\usepackage{comment}
\usepackage{subfig}
\usepackage{color}
\usepackage{graphicx}
\usepackage{mycaption}
\usepackage[figuresright]{rotating}
\usepackage[hyperfootnotes=false,linktocpage=true,breaklinks=true]{hyperref}%the option is important for the toc and ref where otherwise I go out of the page!!!

\newcommand{\ppb}{\ensuremath{p\bar{p}}}
\newcommand{\Et}{\ensuremath{E_T}}
\newcommand{\pt}{\ensuremath{p_T}}

\newcommand{\ba}{\begin{array}}
\newcommand{\ea}{\end{array}}
\newcommand{\bc}{\begin{center}}
\newcommand{\ec}{\end{center}}
\newcommand{\bn}{\begin{enumerate}}
\newcommand{\en}{\end{enumerate}}
\newcommand{\bq}{\begin{equation}}
\newcommand{\eq}{\end{equation}}
\newcommand{\bi}{\begin{itemize}}
\newcommand{\ei}{\end{itemize}}   
\newcommand{\bh}{\begin{math}}
\newcommand{\eh}{\end{math}}
\newcommand{\br}{\begin{flushright}}
\newcommand{\er}{\end{flushright}}
\newcommand{\bl}{\begin{flushleft}}
\newcommand{\el}{\end{flushleft}}
\newcommand{\bt}{\begin{tabular}}
\newcommand{\et}{\end{tabular}}
\newcommand{\flame}

\newcommand{\met}{\mbox{${\not}{E_T}$}}
\newcommand{\metraw}{\mbox{${\not}{E_T}^{\textrm{raw}}$}}

\begin{document}

\ifthenelse{\equal{\LineNumbers}{true}}{
 \pagewiselinenumbers
 \linenumbers
}

\title{Measurement of the $WW$ and $WZ$ production cross section using final states with a charged lepton and heavy-flavor jets in the full CDF Run II data set}

\affiliation{Institute of Physics, Academia Sinica, Taipei, Taiwan 11529, Republic of China}
\affiliation{Argonne National Laboratory, Argonne, Illinois 60439, USA}
\affiliation{University of Athens, 157 71 Athens, Greece}
\affiliation{Institut de Fisica d'Altes Energies, ICREA, Universitat Autonoma de Barcelona, E-08193, Bellaterra (Barcelona), Spain}
\affiliation{Baylor University, Waco, Texas 76798, USA}
\affiliation{Istituto Nazionale di Fisica Nucleare Bologna, $^{ee}$University of Bologna, I-40127 Bologna, Italy}
\affiliation{University of California, Davis, Davis, California 95616, USA}
\affiliation{University of California, Los Angeles, Los Angeles, California 90024, USA}
\affiliation{Instituto de Fisica de Cantabria, CSIC-University of Cantabria, 39005 Santander, Spain}
\affiliation{Carnegie Mellon University, Pittsburgh, Pennsylvania 15213, USA}
\affiliation{Enrico Fermi Institute, University of Chicago, Chicago, Illinois 60637, USA}
\affiliation{Comenius University, 842 48 Bratislava, Slovakia; Institute of Experimental Physics, 040 01 Kosice, Slovakia}
\affiliation{Joint Institute for Nuclear Research, RU-141980 Dubna, Russia}
\affiliation{Duke University, Durham, North Carolina 27708, USA}
\affiliation{Fermi National Accelerator Laboratory, Batavia, Illinois 60510, USA}
\affiliation{University of Florida, Gainesville, Florida 32611, USA}
\affiliation{Laboratori Nazionali di Frascati, Istituto Nazionale di Fisica Nucleare, I-00044 Frascati, Italy}
\affiliation{University of Geneva, CH-1211 Geneva 4, Switzerland}
\affiliation{Glasgow University, Glasgow G12 8QQ, United Kingdom}
\affiliation{Harvard University, Cambridge, Massachusetts 02138, USA}
\affiliation{Division of High Energy Physics, Department of Physics, University of Helsinki and Helsinki Institute of Physics, FIN-00014, Helsinki, Finland}
\affiliation{University of Illinois, Urbana, Illinois 61801, USA}
\affiliation{The Johns Hopkins University, Baltimore, Maryland 21218, USA}
\affiliation{Institut f\"{u}r Experimentelle Kernphysik, Karlsruhe Institute of Technology, D-76131 Karlsruhe, Germany}
\affiliation{Center for High Energy Physics: Kyungpook National University, Daegu 702-701, Korea; Seoul National University, Seoul 151-742, Korea; Sungkyunkwan University, Suwon 440-746, Korea; Korea Institute of Science and Technology Information, Daejeon 305-806, Korea; Chonnam National University, Gwangju 500-757, Korea; Chonbuk National University, Jeonju 561-756, Korea; Ewha Womans University, Seoul, 120-750, Korea}
\affiliation{Ernest Orlando Lawrence Berkeley National Laboratory, Berkeley, California 94720, USA}
\affiliation{University of Liverpool, Liverpool L69 7ZE, United Kingdom}
\affiliation{University College London, London WC1E 6BT, United Kingdom}
\affiliation{Centro de Investigaciones Energeticas Medioambientales y Tecnologicas, E-28040 Madrid, Spain}
\affiliation{Massachusetts Institute of Technology, Cambridge, Massachusetts 02139, USA}
\affiliation{Institute of Particle Physics: McGill University, Montr\'{e}al, Qu\'{e}bec H3A~2T8, Canada; Simon Fraser University, Burnaby, British Columbia V5A~1S6, Canada; University of Toronto, Toronto, Ontario M5S~1A7, Canada; and TRIUMF, Vancouver, British Columbia V6T~2A3, Canada}
\affiliation{University of Michigan, Ann Arbor, Michigan 48109, USA}
\affiliation{Michigan State University, East Lansing, Michigan 48824, USA}
\affiliation{Institution for Theoretical and Experimental Physics, ITEP, Moscow 117259, Russia}
\affiliation{University of New Mexico, Albuquerque, New Mexico 87131, USA}
\affiliation{The Ohio State University, Columbus, Ohio 43210, USA}
\affiliation{Okayama University, Okayama 700-8530, Japan}
\affiliation{Osaka City University, Osaka 588, Japan}
\affiliation{University of Oxford, Oxford OX1 3RH, United Kingdom}
\affiliation{Istituto Nazionale di Fisica Nucleare, Sezione di Padova-Trento, $^{ff}$University of Padova, I-35131 Padova, Italy}
\affiliation{University of Pennsylvania, Philadelphia, Pennsylvania 19104, USA}
\affiliation{Istituto Nazionale di Fisica Nucleare Pisa, $^{gg}$University of Pisa, $^{hh}$University of Siena and $^{ii}$Scuola Normale Superiore, I-56127 Pisa, Italy, $^{mm}$INFN Pavia and University of Pavia, I-27100 Pavia, Italy}
\affiliation{University of Pittsburgh, Pittsburgh, Pennsylvania 15260, USA}
\affiliation{Purdue University, West Lafayette, Indiana 47907, USA}
\affiliation{University of Rochester, Rochester, New York 14627, USA}
\affiliation{The Rockefeller University, New York, New York 10065, USA}
\affiliation{Istituto Nazionale di Fisica Nucleare, Sezione di Roma 1, $^{jj}$Sapienza Universit\`{a} di Roma, I-00185 Roma, Italy}
\affiliation{Texas A\&M University, College Station, Texas 77843, USA}
\affiliation{Istituto Nazionale di Fisica Nucleare Trieste/Udine; $^{nn}$University of Trieste, I-34127 Trieste, Italy; $^{kk}$University of Udine, I-33100 Udine, Italy}
\affiliation{University of Tsukuba, Tsukuba, Ibaraki 305, Japan}
\affiliation{Tufts University, Medford, Massachusetts 02155, USA}
\affiliation{University of Virginia, Charlottesville, Virginia 22906, USA}
\affiliation{Waseda University, Tokyo 169, Japan}
\affiliation{Wayne State University, Detroit, Michigan 48201, USA}
\affiliation{University of Wisconsin, Madison, Wisconsin 53706, USA}
\affiliation{Yale University, New Haven, Connecticut 06520, USA}

\author{T.~Aaltonen}
\affiliation{Division of High Energy Physics, Department of Physics, University of Helsinki and Helsinki Institute of Physics, FIN-00014, Helsinki, Finland}
\author{S.~Amerio}
\affiliation{Istituto Nazionale di Fisica Nucleare, Sezione di Padova-Trento, $^{ff}$University of Padova, I-35131 Padova, Italy}
\author{D.~Amidei}
\affiliation{University of Michigan, Ann Arbor, Michigan 48109, USA}
\author{A.~Anastassov$^x$}
\affiliation{Fermi National Accelerator Laboratory, Batavia, Illinois 60510, USA}
\author{A.~Annovi}
\affiliation{Laboratori Nazionali di Frascati, Istituto Nazionale di Fisica Nucleare, I-00044 Frascati, Italy}
\author{J.~Antos}
\affiliation{Comenius University, 842 48 Bratislava, Slovakia; Institute of Experimental Physics, 040 01 Kosice, Slovakia}
\author{G.~Apollinari}
\affiliation{Fermi National Accelerator Laboratory, Batavia, Illinois 60510, USA}
\author{J.A.~Appel}
\affiliation{Fermi National Accelerator Laboratory, Batavia, Illinois 60510, USA}
\author{T.~Arisawa}
\affiliation{Waseda University, Tokyo 169, Japan}
\author{A.~Artikov}
\affiliation{Joint Institute for Nuclear Research, RU-141980 Dubna, Russia}
% ...

\collaboration{CDF Collaboration\footnote{With visitors from
$^a$Istituto Nazionale di Fisica Nucleare, Sezione di Cagliari, 09042 Monserrato (Cagliari), Italy,
$^b$University of California Irvine, Irvine, CA 92697, USA,
$^c$University of California Santa Barbara, Santa Barbara, CA 93106, USA,
$^d$University of California Santa Cruz, Santa Cruz, CA 95064, USA,
$^e$Institute of Physics, Academy of Sciences of the Czech Republic, 182~21, Czech Republic,
$^f$CERN, CH-1211 Geneva, Switzerland,
$^g$Cornell University, Ithaca, NY 14853, USA,
$^h$University of Cyprus, Nicosia CY-1678, Cyprus,
$^i$Office of Science, U.S. Department of Energy, Washington, DC 20585, USA,
$^j$University College Dublin, Dublin 4, Ireland,
$^k$ETH, 8092 Z\"{u}rich, Switzerland,
$^l$University of Fukui, Fukui City, Fukui Prefecture, Japan 910-0017,
$^m$Universidad Iberoamericana, Lomas de Santa Fe, M\'{e}xico, C.P. 01219, Distrito Federal,
$^n$University of Iowa, Iowa City, IA 52242, USA,
$^o$Kinki University, Higashi-Osaka City, Japan 577-8502,
$^p$Kansas State University, Manhattan, KS 66506, USA,
$^q$Brookhaven National Laboratory, Upton, NY 11973, USA,
$^r$University of Manchester, Manchester M13 9PL, United Kingdom,
$^s$Queen Mary, University of London, London, E1 4NS, United Kingdom,
$^t$University of Melbourne, Victoria 3010, Australia,
$^u$Muons, Inc., Batavia, IL 60510, USA,
$^v$Nagasaki Institute of Applied Science, Nagasaki 851-0193, Japan,
$^w$National Research Nuclear University, Moscow 115409, Russia,
$^x$Northwestern University, Evanston, IL 60208, USA,
$^y$University of Notre Dame, Notre Dame, IN 46556, USA,
$^z$Universidad de Oviedo, E-33007 Oviedo, Spain,
$^{aa}$CNRS-IN2P3, Paris, F-75205 France,
$^{bb}$Texas Tech University, Lubbock, TX 79609, USA,
$^{cc}$Universidad Tecnica Federico Santa Maria, 110v Valparaiso, Chile,
$^{dd}$Yarmouk University, Irbid 211-63, Jordan,
$^{ll}$Universite catholique de Louvain, 1348 Louvain-La-Neuve, Belgium,
$^{oo}$University of Z\"{u}rich, 8006 Z\"{u}rich, Switzerland,
$^{pp}$Massachusetts General Hospital and Harvard Medical School, Boston, MA 02114 USA,
$^{qq}$Hampton University, Hampton, VA 23668, USA,
$^{rr}$Los Alamos National Laboratory, Los Alamos, NM 87544, USA
}}
\noaffiliation

\date{\today}

\begin{abstract}
We present a measurement of the total {\it WW} and {\it WZ} production cross sections in $p\bar{p}$ 
collision at $\sqrt{s}=1.96$ TeV, in
a final state consistent with leptonic $W$ boson decay and jets originating from heavy-flavor quarks 
from either a $W$ or a $Z$ boson decay.
This analysis uses the full data set collected with the CDF II detector during Run II of the
Tevatron collider, corresponding to an integrated luminosity of 9.4 fb$^{-1}$. 
An analysis of the dijet mass spectrum 
provides $3.7\sigma$ evidence of the summed production processes of either {\it WW} or {\it WZ} bosons with a
measured total cross section of 
$\sigma_{WW+WZ} = 13.7\pm 3.9$~pb. 
Independent measurements of the {\it WW} and {\it WZ} production cross sections are allowed by the 
different heavy-flavor decay-patterns of
the $W$ and $Z$ bosons 
and by the analysis of secondary-decay vertices reconstructed within heavy-flavor jets. 
The productions of {\it WW} and of {\it WZ} dibosons are independently seen with significances of 
$2.9\sigma$ and $2.1\sigma$,
respectively, with total cross sections of  $\sigma_{WW}=  9.4\pm 4.2$~pb and 
$\sigma_{WZ}=3.7^{+2.5}_{-2.2}$~pb. 
The measurements are consistent with standard-model predictions.
\end{abstract}

\maketitle

%==================================================================
\section{Introduction}

The production of a pair of $W$ or $Z$ vector bosons is a process of 
primary interest  at hadron colliders.
Measurements of the production of different vector-boson pairs
probe the multiple gauge-boson 
couplings~\cite{TGC_Degrande201321} predicted by the standard model (SM),
and provide
a benchmark for analyses designed to study lower-cross-section processes
 sharing the same final states, like Higgs-boson production.

Diboson production has been extensively studied  by the CDF, D0, ATLAS and CMS collaborations
using final states where both bosons decay leptonically.
The multiplicity of charged leptons and neutrinos allows for the separation 
of {\it WW}, {\it WZ}, and {\it ZZ} production 
with the advantage of a distinctive experimental signature.  
Current measurements of {\it WW}~\cite{ATLAS_WW_PhysRevD.87.112001, ATLAS_WW_8TeV, CMS_WW2015, CDF_WWj_PhysRevD.91.111101}, 
{\it WZ}~\cite{ATLAS_WZ_2012, ATLAS_WZ_2016_8TeV, D0_WZZZlep_2012}, and {\it ZZ}~\cite{ATLAS_ZZ_llll, ATLAS_ZZ_llll_13TeV, CMS_ZZ_llll, CDF_ZZ_llll} production cross sections
have fractional precision in the $4$\%$-25$\% range.

The analysis of final states with one of the two bosons decaying leptonically and the 
other hadronically (hereafter called the semileptonic final state)
is more challenging  because of the large background from QCD 
hadron production and  the poor resolution in the reconstructed energy of hadronic jets compared  to charged leptons.
At the Tevatron and Large Hadron Collider (LHC) experiments, 
the measurements of the combined {\it WW+WZ} production cross section for the semileptonic final
states have precisions of
$15$\%$-30$\%~\cite{CDF_DibosonMjj_lvjj, CDF_DibosonME_lvjj,D0_Diboson_lvjj, ATLAS_Diboson_lvjj, CMS_Diboson_lvjj}.
The separate measurement of {\it WW} and {\it WZ} production is highly 
challenging
 because the width of the dijet-mass distribution is larger than the mass 
difference between the $Z$ and $W$ bosons. 
One method to distinguish the  two production modes is to use the 
different heavy flavor (HF) 
hadronic decays involving $W$ or $Z$ bosons, namely $W\to cs$, and $Z\to c\bar{c}$ or  $Z\to b\bar{b}$. 
The D0 collaboration used an experimental signature targeting $Z\to b\bar{b}$ 
decays to measure {\it WZ} production in the semileptonic final state with an uncertainty 
of $100$\%$-120$\%~\cite{D0_Diboson_lvjj}. A precise measurement of the 
{\it WZ} 
production cross-section in this final state is still missing.

In searches for the as-yet-unobserved decay of the SM Higgs bosons to a pair of $b$ quarks,  Tevatron~\cite{Tevatron_Hbb} and LHC~\cite{ATLAS_VHbb2015,CMS_VHbb2014} 
experiments obtained the highest sensitivity by investigating the {\it WH} and {\it ZH} production modes. Because of the small expected signal yield and of the large 
backgrounds,  multivariate discriminating algorithms have been used extensively and searches in the final states with zero, one, and two leptons have been combined together. 
In the same analyses, the total {\it WZ+ZZ} production cross section has been measured in the semileptonic final states enriched in HF hadrons with 
an uncertainty of approximately $20\%$. 

The goal of the analysis described in this paper is the measurement of 
the combined and separate  {\it WW} and 
{\it WZ} production cross-sections using the differences in heavy flavour  
(HF) hadronic decays involving 
the $W$ and $Z$ bosons. 
The analysis is also a benchmark for the  CDF search for 
{\it WH} production~\cite{WHbb_2012}, the most sensitive analysis channel contributing to searches
for the SM Higgs boson decaying to the $b\bar{b}$ final state at the 
Tevatron~\cite{Tevatron_Hbb}.

This measurement is based on the full proton-antiproton ($p\bar{p}$) collision data set collected with the 
CDF II detector at the Tevatron collider, corresponding to an 
integrated luminosity of 9.4~fb$^{-1}$  at a center-of-mass energy of  
$\sqrt{s}=1.96$ TeV. 

Events are selected by requiring only one fully-reconstructed electron or muon candidate  $\ell$
and an imbalance in the total energy measured with respect to the plane transverse to the colliding beams (\met), 
indicative of the presence of a neutrino. Both requirements strongly suppress the  background  of {\it ZZ} production where one of the two bosons decays hadronically and the other leptonically.
A support-vector-machine (SVM) algorithm~\cite{NIM_SVM} is used to select events 
consistent with $W\to \ell\nu$+jets production. Finally, the hadronic decay 
of the $W$ or $Z$ boson is identified by requiring events with two jets with a large component of the momentum transverse to
the beam (transverse momentum) and where, in at least one of them, a 
secondary-decay vertex is identified in at least one jet, indicating
the presence of a $b$ or $c$ hadron (HF tag). The resonant $W$ or 
$Z$ boson 
signal is separated from the large nonresonant background by studying the 
dijet mass spectrum. In combination with this, a 
flavor-separator neural network~\cite{single_top} (flavor-separator NN) is 
used to separate jets originating from a charm or bottom quark, enhancing 
the sensitivity
to {\it WW} or {\it WZ} production. 
The method with which the {\it WW} and {\it WZ} contributions are disentangled is
novel.
A Bayesian statistical analysis is then 
 used to extract the signal cross sections by comparing the data to the background predictions.
Further details on the analysis are in Ref.~\cite{WZThesis}.

The paper is organized as follows: the CDF II detector is briefly described in Sec.~\ref{sec:detector}; the selection of signal candidates is
reported in Sec.~\ref{sec:selection}; the details of the Monte Carlo~(MC) 
simulation used in the analysis are given in Sec.~\ref{sec:MC}; the background estimation is
described in Sec.~\ref{sec:Back}. The signal-to-background discrimination is discussed in Sec.~\ref{sec:discriminant}, before the description of the  signal-extraction statistical-analysis and its results, in Sec.~\ref{sec:fit}. Conclusions are given in Sec.~\ref{sec:conclusions}.

\section{The CDF experiment}\label{sec:detector}

The CDF II detector, described in detail in Ref.~\cite{top_lj2005}, 
operated at the Tevatron collider from 2001 until 2011. 
It  was a multipurpose particle detector
composed of a charged-particle tracking system  immersed in a $1.4$ T  
axial magnetic field and surrounded by calorimeters and muon chambers. 
The detector had azimuthal symmetry around the beam axis and forward-backward symmetry with respect to 
the collision point. Particle trajectory (track) coordinates are described 
in a cylindrical-coordinate system
with the $z$-axis along the proton beam, the azimuthal angle $\phi$ about the beam axis,
and the polar angle $\theta$, measured with respect to the proton beam direction. The following variables are defined: pseudorapidity, $\eta = - \ln(\tan(\theta/2))$;
transverse energy, $\Et = E \sin\theta$;  transverse momentum, 
$p_T = p \sin \theta$; and angular distance between two particles $A$ and $B$ as $\Delta R \equiv \sqrt{(\phi_{A} -\phi_{B}) ^2 + (\eta_{A} - \eta_{B})^2 }$.

The charged-particle tracker was composed by a set of silicon detectors~\cite{svx, isl} and by an open-cell drift chamber~\cite{cot} covering radial ranges up to about 30 cm and 140 cm, respectively.
The inner silicon detectors covered the pseudorapidity range 
$|\eta| < 2$, provided a spatial resolution on each measurement point 
in the $r - \phi$ plane of approximately $11$~$\mu$m, and they achieved 
a resolution on the track transverse impact parameter~\cite{d0}, 
$\sigma(d_0)$ of 
about $40$ $\mu$m, of which about 30~$\mu$m was due to the transverse size of the Tevatron beam.
The $3.1$~m long drift chamber, 
 covering the region of $|\eta| < 1.0$, ensured a measurement of charged-particle transverse momenta
with resolution of $\sigma_{p_T}/p_T \approx 0.07\%\cdot p_T$, where momenta are in units of  GeV/$c$.

 The tracking system was surrounded by calorimeters,
which measured the energies and the directions of electrons, photons, and
jets of hadronic particles. The electromagnetic calorimeters used lead and
scintillating-tile sampling technology,
while the hadronic calorimeters were composed of scintillating tiles with steel absorber. 
The calorimeters were divided into central and plug sections, each segmented in  $\eta - \phi$ projective-geometry towers pointing towards the nominal interaction point.
The central section, composed of the central electromagnetic~\cite{Balka:1987ty} and central and end-wall hadronic calorimeters~\cite{Bertolucci:1987zn}, covered the region $|\eta|< 1.1$. The end-plug electromagnetic~\cite{Albrow:2001jw} and end-plug hadronic calorimeters extended the coverage to $|\eta| < 3.6$.  At a depth of about six radiation lengths inside the 
electromagnetic calorimeters, detectors with finer $\eta - \phi$ segmentation~\cite{Apollinari:1998bg} were 
used to provide position measurement and shape information for electromagnetic showers.

In the outermost layer of the CDF II detector, a composite set of planar 
multiwire-drift chambers was
used for muon identification~\cite{Ascoli:1987av,Artikov:2005}.  The detectors were arranged into 
the central-muon section covering the region $|\eta|<0.6$, 
the central-muon extension covering the region $0.6<|\eta|<1.0$, and the barrel-muon chambers covering the region 
$1.0<|\eta|<1.5$.

The imbalance in the magnitude of the vector sum of all calorimeter-tower-energy 
depositions projected on the transverse plane, dubbed raw missing transverse energy or \metraw, is used to infer 
the transverse momentum of additional particles escaping detection, most notably neutrinos.

A set of gaseous Cherenkov counters~\cite{clc_performance} located at large pseudo-rapidity, 
\mbox{$3.6< |\eta| <4.6$}, was used to measure the instantaneous luminosity through the rate of 
$p\bar{p}$ inelastic collisions.

During data taking, collision events were selected in real time to be recorded on tape by a three-level filtering system (trigger)~\cite{trigger1, trigger2}
that used a combination of multiple selection criteria, 
called trigger paths, in order to reduce the initial event rate of 1.7 MHz to about 100 Hz.

\section{Data Sample and event selection}\label{sec:selection}

We maximize the acceptance of events containing one electron or muon,
missing transverse energy, and two heavy-flavor jets in the final state.

The trigger selection and lepton and jet identifications follow those developed in searches for rare
processes, such as {\it WH}$\to\ell\nu+b\bar{b}$~\cite{WHbb_2012} and
 $s$-channel-single-top quark production~\cite{signletop_ljet_2014}. Novel 
strategies are used in the selection of the heavy-flavor 
jet-enriched sample and for the rejection of  background 
events originating from multijet production.

\subsection{Trigger selection and categorization}\label{sec:OnlineSelection}

Data were collected using several trigger paths, which are categorized by the following four analysis regions, homogeneous in kinematic and background composition:
\begin{enumerate}
\item[(1)]{\em Central electrons.} Collected by requiring a track from a charged particle with $\pt>18$~GeV/$c$ matched to an 
electromagnetic cluster with $\Et>18$~GeV and  $|\eta|<1.0$. 
\item[(2)]{\em Central muons.} Collected by requiring a track from a charged particle with 
$p_T>18$~GeV/$c$ and $|\eta|<1.0$
matched to hits detected in a central muon chamber.
\item[(3)]{\em Forward electrons.} Collected by requiring clusters in the electromagnetic calorimeter, reconstructed in the region $1.2 < |\eta| < 2.0$ and with $\Et> 20$~GeV, and missing transverse energy, reconstructed with \metraw$>15$~GeV.
\item[(4)]{\em Extended muons.} 
As detailed in Ref.~\cite{WH7.5}, a combination of trigger paths able to collect events containing a neutrino and high-\pt\, jets is used
in order to recover events that, at trigger level, are not identified as containing a charged lepton.
This data set is
referred to as ``extended muons'', since large \metraw\ may be originating from
 high-\pt\ muons depositing a small amount of energy in the calorimeter system. 
\end{enumerate}
Due to the changes in instantaneous luminosity that occurred over the ten years of detector operation, 
 some trigger selections were modified or their rates were
 decreased by randomly accepting a fixed fraction of the
events that met the trigger selection. 
Trigger efficiencies were measured on data as functions of the
instantaneous luminosity and kinematic properties of the events~\cite{NIM_Adrian}.
For the triggers used for the extended-muon category, the efficiency
is parametrized as a function of the missing transverse energy reconstructed using only 
calorimeter information, i.e., without accounting for any correction due to the momentum of 
detected muons.
The uncertainty on the total trigger efficiencies ranges 
from 1\% (central muons) to 3\% (extended muons).

\subsection{$W\to\ell\nu$ plus HF-jets-candidate selection}\label{sec:ev_sel}

Candidate signal events are selected off line to be consistent with the 
production of a $W$ boson decaying leptonically 
and of a second particle decaying to two jets containing HF hadrons.

As a first step, events with exactly one muon (electron) candidate of \pt$>20$ GeV/$c$ (\Et$>20$ GeV) are selected.
Muon and electron candidates are required to originate from a primary vertex ~\cite{vtx} within $\pm$ 60~cm of the center
of the CDF II detector, as measured along the beam line.
A total of ten lepton-reconstruction algorithms, encompassing four 
``tight'', five ``loose'', and one ``track-only'' classes are used.
Two classes of tight-electron candidates are identified, associated with 
the central ($|\eta| < 1.0$) and forward ($1.2 < |\eta| < 2.0$) 
regions of the electromagnetic calorimeter, using selection requirements based on five  electromagnetic-shower-shape profiles and calorimeter variables~\cite{WZThesis}.
Two classes of tight-muon candidates are selected in the central muon subdetectors, for $|\eta| < 0.6$ and  $0.6 < |\eta| < 1.0$. 
A set of five loose-muon-class identification criteria is defined to recover signal acceptance up to $|\eta| < 1.4$ and in the gaps of the muon subdetectors.
Each lepton-candidate class requires the presence of  a reconstructed track in the drift chamber, 
except for the forward-electron class where only partial track-reconstruction in the silicon 
detectors is required~\cite{phx_alg}.
The track-only class consists of good-quality tracks reconstructed in the drift chamber with $\Delta R>0.4$ from any reconstructed hadronic jet, 
as defined in the following, with \Et$>20$ GeV. About 15\% of the track-only lepton candidates are 
expected to originate from electrons or $\tau$ lepton hadronic decays.
To increase the purity of prompt leptons from $W$-boson decays, lepton candidates are required to be well isolated.
Calorimeter isolation is used for tight and loose lepton classes: The energy reconstructed in a cone of radius
$\Delta  R=0.1$ around the lepton must be less than 10\% of the equivalent \pt\, (\Et) of the muon (electron) candidate. 
Isolation in the tracking volume is used for the track-only leptons. The transverse momentum 
associated with the sum of all the tracks 
inside a cone of radius $\Delta R=0.1$ around the track-only lepton candidate must be less than 10\% of the \pt\, of the candidate. 

As a second step, events with two hadronic jets of \Et$>20$ GeV and $|\eta| < 2.0$ are selected.
Jets are identified and reconstructed from clusters of calorimeter
energies contained within a cone of radius $\Delta R=0.4$~\cite{jetclu}.  The calorimeter towers corresponding
to the energy deposit of any tight-electron candidate are excluded from the jet-reconstruction 
algorithm.  The \Et\, of a jet, in experimental and simulated data, 
is calculated from the sum of the calorimeter clusters~\cite{jet_corr1} and undergoes the following 
jet-energy scale  calibration procedure to reproduce more accurately the energy of the originating hadrons: 
calorimeter response is adjusted to be independent of $\eta$, energy contributions from multiple $\ppb$ 
interactions are removed based on the number of reconstructed interaction vertices
per bunch crossing in each event, and nonlinearities in the calorimeter response are corrected.

Finally, events with $W\to\ell\nu$ candidates are required to be consistent with the presence of a neutrino. 
The \metraw\, is corrected for the momentum of the muon and track-only candidates, and for the jet-energy-scale calibration 
of the jets with \Et$> 12$ GeV and $|\eta|< 2.4$. Unlike the case for generic jet-energy-scale calibration, described above, 
effects due to multiple interactions are not taken into account as high-energy neutrinos are assumed to originate only from the primary $\ppb$ interaction. 
The value of \metraw\, after these corrections, \met\,,  is required to be greater than 15 GeV.

The selected data sample is still rich in multijet (MJ) events produced by QCD processes erroneously reconstructed as having the $W\to\ell\nu$ decay signature. 
A MJ suppression algorithm, based on a SVM discriminant, 
described in Sec.~\ref{sec:multijet_sel}, is used to improve the signal-to-background ratio.

The selection criteria described are referred to as the ``pretag'' selection. The selected sample 
consists of $232\,145$ events, most of them containing a $W$ boson
decaying leptonically and produced in association with jets originating from light-flavor quarks 
and gluons,  as no jet-flavor discrimination is performed. 
%The pretag sample is used as a control sample in the analysis of the HF-enriched sample. 
A subset of data enriched in events with jets originating from HF quarks is
obtained through the identification of 
long-lived hadrons that decay away from the primary vertex.
The \textsc{secvtx}~\cite{top_lj2005} algorithm attempts to reconstruct a secondary vertex 
in each jet containing at least two charged particles of \pt$>0.5$ GeV$/c$ (``taggable'' jet). If a vertex is found
and is significantly displaced from the primary \ppb\, interaction, then the jet is ``tagged'' as a HF jet. 
The algorithm is operated with different (``tight'' or ``loose'') requirements on the tracks and vertex-reconstruction quality.
Events are classified in distinct signal regions depending on the number of HF-tagged jets:
\begin{enumerate}
\item[(1)]{\em Single-tag (or one-tag).} Events with one tight \textsc{secvtx}-tagged jet are classified in this signal region.
The signal selection efficiency of this category, which also includes the events where the additional jet may have a loose-\textsc{secvtx} tag, is relatively high.
  The identification efficiency for {\it WZ} events containing one jet originating from a $b$ quark is approximately 42\%, 
  while the identification efficiencies for {\it WZ} or {\it WW} events with one jet originating from a $c$ quark are about 12\% and 8\%, respectively. 
\item[(2)]{\em Double-tag (or two-tag).} Events with two jets tagged both by the tight or both by the loose \textsc{secvtx} algorithm are classified in 
this signal region, with the double-loose tag category accounting for approximately 10\% of the total double-tag-data.
The reconstruction efficiency for {\it WZ} events with the $Z$ boson decaying to $b$ quarks in this category is about 11\%, with additional small contributions 
to the signal yield from $Z\to c\bar{c}$ decays. 
\end{enumerate}

\subsection{Suppression of multijet background using an SVM discriminant}\label{sec:multijet_sel}

The use of loose-lepton-identification algorithms and of a low-\met\, requirement is 
well suited to recover signal acceptance in searches for processes with $W$-boson leptonic decays and small expected yield.
However, this enhances the  background contribution from QCD multijet events classified as having 
a $W$-boson-like signature, e.g., if a particle in a jet 
meets the lepton identification criteria and moderate \met\, is generated from energy 
mismeasurement. 
To cope with this, a multivariate multijet rejection strategy based on a support-vector-machine algorithm is developed~\cite{NIM_SVM}.

A discussion of the SVM algorithm and its usage is available in Ref.~\cite{BShop_machine_learning}; 
however, the basic concepts are briefly illustrated in the following: The SVM algorithm builds the best separating hyper-plane, 
called the ``margin'', between two classes of events, the signal and the background training sets, with each event
represented as a point associated with a vector of coordinates in the multidimensional space of the discriminating variables.
If the two sets of vectors are not linearly separable, as is often the case in real applications, an appropriate transformation function, the ``kernel'', 
is used to map the vectors into a higher-dimension space where linear separation becomes possible. 
The margin is completely defined by a relatively small subset of the input training points, called ``support vectors''.
A new point is classified according to its SVM output value, which is the signed distance of the 
point with respect to the margin in the space where linear separation is obtained. A negative or positive SVM output value corresponds to background-like or 
signal-like classification, respectively.

The specific choices used in this analysis are described in more detail in Ref.~\cite{NIM_SVM}.
The input training sets used are simulated  $W\to e\nu+$jets signal events and MJ-background events from data, as described in Sec.~\ref{sec:back_pretag}. 
Each training point is described by a set of kinematic variables characterizing the $W\to\ell\nu$ decay,
the measured leading jet momentum,  \met\, and \metraw\,, as well as the angles between the reconstructed objects.
The two sets of vectors from the signal and background training sets are not linearly separable;
therefore, a kernel transformation is used.
To deal with possible mismodeling and biases in the training models, feedback from a data control sample, selected with loose $W+$jets requirements, is included 
in the training procedure. This ensures that the SVM variable evaluated in simulated events closely reproduces the results 
obtained in data.

Two different SVM discriminants are developed, one for events with the lepton reconstructed in the 
central ($|\eta|<1.1$) 
region of the detector and one for the forward ($1.1<|\eta|<2.0$).
The event selection criteria are therefore defined by a threshold on the SVM output value, as shown in Fig.~\ref{fig:svm_pretag} where higher 
SVM thresholds are used in cases where the background is larger to obtain optimum performance.
Events in the  central-electron and extended-muon categories are required to have an output value 
of the 
central-SVM discriminant greater than zero.
This selection rejects about $90$\% of the MJ 
background, as measured on the training sample,  with a signal selection efficiency of $95$\% for {\it WW} or {\it WZ} simulated events. A relaxed  
selection threshold of \mbox{SVM output greater than $-0.5$} is used for events in the central-muon category; 
this yields approximately 98\% signal efficiency while still rejecting about 90\% of the MJ events.
Events  in the forward-electron category, approximately 12\% of the total sample, 
are required to have an output value of the forward SVM discriminant greater than one; this rejects about $90$\% of the MJ 
background with a signal-selection efficiency of approximately $82$\% for simulated {\it WW} or {\it WZ} events.

\section{Event Simulation}\label{sec:MC}

Several MC event generators are used to model background and signal
processes. The \textsc{pythia}~v6.2~\cite{pythia} event generator at leading order
(LO) in the strong-interaction coupling is used for the simulation of diboson {\it WW} and {\it WZ} signal, production and decay in inclusive final states. The same generator is used to simulate the 
contribution from the {\it ZZ} production and the production of a 125 GeV$/c^2$ Higgs boson in association with a $W$ or $Z$ boson, in the decay channels $W\to\ell\nu$, $Z\to\ell\ell/\nu\nu$, and $H\to b\bar{b}$.

The \textsc{powheg}~\cite{powheg} generator at next-to-leading order (NLO) in the
strong-interaction coupling  
 is used for single- and pair-production of top-quarks with 172.5 GeV/$c^2$ mass. 
The \textsc{alpgen}~\cite{alpgen} LO generator is used for the matrix-element calculations 
of $Z(\ell\ell)$\,+\,jets and $W(\ell\nu)$\,+\,jets processes, resulting in samples with up to four  partons in the final state. 
The \textsc{alpgen}~\cite{alpgen} LO generator is also used for the simulation of $W(\ell\nu)$\,+\,HF-partons samples with 
massive HF quarks ($b\bar{b}$, $c\bar{c}$, single $c$) accounted for in the matrix-element calculation.
For all the simulated processes, the  \textsc{pythia}~v6.2 program is used for the simulation of parton-showers and hadronization. 

A generated-to-reconstructed jet-matching scheme~\cite{alpgen} is used to correctly account for 
jets produced by the matrix element and the parton showers 
within each of the samples simulated with \textsc{alpgen}. As separate samples with $W(\ell\nu)$\,+\,HF partons are used,  
care is taken to avoid double counting of the  HF-parton phase space simulated by the 
matrix element or by the parton showers 
in different samples, as collinear QCD radiation is accounted for by the parton-shower program.
This is done with an overlap-removal scheme based on jet cones of radius $\Delta R=0.4$. 
For each generated event, if  two HF partons originating from the matrix element are found to be in the same jet cone, 
the event is discarded. An event is also discarded if two HF partons 
originating from parton showers are reconstructed in two separate jet cones. 
This procedure allows for a smooth transition between the small $\Delta R$ region, where the dynamics is dominated by
the parton shower, and the large $\Delta R$ region, where the hard scattering dominates.

In addition to the jet-energy scale calibration procedure described in Section~\ref{sec:ev_sel}, 
a specific corrections~\cite{dijet_marco} is applied on simulated jets to account for their quark-like or 
gluon-like nature.

The distributions of the longitudinal 
momenta of the different types of quarks and gluons within the proton, parametrized as functions of the momentum transfer of 
the collision, are given by parton distribution functions (PDFs). The CTEQ5L~\cite{cteq5l} PDFs are used in 
generating all MC samples in this analysis.
The underlying event model~\cite{UE_explain} is tuned to data~\cite{CDF_UnderlingEvent}.
The simulated events are processed through a \textsc{geant3}-based detector 
simulation~\cite{CDF_SIMULATION} and then 
through the same reconstruction software used for experimental data.

\begin{table*}[t!]
  \centering
  \caption{Summary of the simulated signal and background processes (first column) together with the corresponding MC programs used for the event generation (second column) and with the normalization cross sections used, when available (third column).}\label{tab:cx}
  \begin{tabular}{c c c}
    \hline\hline
    Process &  Generator & Cross section [pb] \\
    \hline
    {\it WW}  &  \textsc{pythia}  & $11.34 \pm 0.66$, NLO~\cite{MCFM}\\ 
    {\it WZ}  &  \textsc{pythia}  & $3.47  \pm 0.21$, NLO~\cite{MCFM}\\
    {\it ZZ}  &  \textsc{pythia}  & $3.62  \pm 0.22$, NLO~\cite{MCFM}\\
    $W(\ell\nu)H(b\bar{b})$  &  \textsc{pythia} & $(2.4\pm0.2$)$\times 10^{-2}$, NNLO~\cite{HiggsCx}\\
    $Z(\ell\ell/\nu\nu) H(b\bar{b})$ &  \textsc{pythia} & $(4.6\pm 0.5$)$\times 10^{-3}$, NNLO~\cite{HiggsCx}\\
    $t\bar{t}$ & \textsc{powheg}  & $7.04 \pm 0.49$, NNLO~\cite{ttbarCx}\\
    single-top-$s$ channel &  \textsc{powheg}  &  $1.04 \pm 0.07$, NNLO~\cite{stopCx}\\
    single-top-$t$ channel &   \textsc{powheg} &  $2.10 \pm 0.19$, NNLO~\cite{stopCx}\\
    $Z(\ell\ell)$ + jets &  \textsc{alpgen} & $787 \pm 85$, measurement~\cite{zjetsCx}\\
    $W(\ell\nu)$ + jets & \textsc{alpgen} &  Based on data \\
    \hline\hline
  \end{tabular}
\end{table*}

The expected event yield of a simulated process is calculated as
\begin{equation}\label{eq:mc_norm}
N = \sigma  \mathcal{L} A\textrm{,}
\end{equation}
where, for each process, $\sigma$ is the NLO (or NNLO) cross section reported in Table~\ref{tab:cx},  
$A$ is the acceptance for each lepton and HF-tag category, and $\mathcal{L}$
is the integrated luminosity collected by the appropriate trigger path.

The acceptance $A$ is estimated from simulated events and corrected for lepton reconstruction and 
HF tagging efficiencies 
observed in data by means of scale factors,
\begin{equation}\label{lepSF}
  SF =\frac{ \epsilon^{\rm MC}}{\epsilon^{\rm data}}\textrm{,}
\end{equation}
where $\epsilon^{\rm MC}$ is the efficiency associated with the lepton reconstruction or the HF-tagging, determined from simulation,
 and $\epsilon^{\rm data}$ is the corresponding efficiency measured in a data control sample.

The scale factors corresponding to each lepton identification algorithm are measured in $Z\to\ell\ell$ events and then applied 
as weights to each simulated event entering in the evaluation of $A$ in Eq.~(\ref{eq:mc_norm}).  
The $SF$s corresponding to the
tight lepton identification algorithms are known with a relative uncertainty below one percent. A relative uncertainty of 
approximately 2\% is derived for the loose-muon identification algorithms, with the main data-MC 
differences arising in the 
modeling of the isolation and of the muon-chamber response.  
A relative uncertainty of 4.5\%~\cite{WH7.5} is derived for the isolated-track category, 
primarily to account for the flavor of the particle associated with the track.

The per-jet simulated HF-tagging efficiencies are corrected with scale factors derived from HF-enriched MJ samples~\cite{top_lj2005}.  
The per-jet corrections are then propagated to the events entering in the computation of $A$ in Eq.~(\ref{eq:mc_norm}) 
through the following formulas:
\begin{equation}\label{eq:tag}
  \omega_{\textrm{1-tag}}= SF_{\textrm{j1}}\big( 1 - SF_{\textrm{j2}}\big) +SF_{\textrm{j2}} \big(1 - SF_{\textrm{j1}}\big) 
\end{equation}
\begin{equation}\label{eq:tag2}
   \omega_{\textrm{2-tags}}= SF_{\textrm{j1}}\cdot SF_{\textrm{j2}} \textrm{,}
\end{equation}
where $SF_{\textrm{j1}}$, $SF_{\textrm{j2}}$ are the HF-tagging-efficiency corrections relative to the first and second jet in the event 
and $\omega_{\textrm{1-tag}}$,  $\omega_{\textrm{2-tags}}$ are the probabilities of each simulated event to meet the one-tag or
 the two-tag selection, respectively.
The $SF$s of the tight and loose \textsc{secvtx} criteria are close to unity and do not show 
$\eta$ or \pt\ dependences in the kinematic range of interest. 
The relative systematic uncertainty associated with the tight (loose) working point $SF$ is  5\% (7\%) for each jet originating 
from $b$ quarks and about twice those values for jets originating from $c$ quarks.

\section{Background estimate}\label{sec:Back}

The selected events are likely to originate from processes characterized by the $\ell\nu\,+\,$HF signature. However, in addition to the signal, several other 
processes  are classified as having the same final state:
\begin{enumerate}
\item[(1)]{\em W+HF.} Processes ($W + b\bar{b}$, $W + c\bar{c}$ and $W + c$) involving the production of a $W$ boson in association with HF quarks,
mainly from radiated gluons. 
This category represents the main source of irreducible background.
\item[(2)]{\em W+LF.} Events with a real $W$ boson produced in association with one or more light-flavor (LF) jets mistakenly identified 
as a HF jet by the
\textsc{secvtx} algorithm. Mistags are generated because of the finite resolution of the tracking detectors, material interactions,
or from long-lived-LF hadrons ($\Lambda$  and $K^0_s$) that produce real displaced vertices.
\item[(3)]{\em EWK.} Background contributions from processes with a real lepton and HF jets. 
They originate from single-top quark and top-quark pair production, production of $Z$ boson plus jets, and, 
to a lesser extent, {\it WH} and {\it ZZ} production. 
\item[(4)]{\em MJ.} Contributions from QCD  multijet production giving a $W$-boson-like signature when 
one of the particles in a jet passes the lepton identification criteria and energy mismeasurement in the event results in false \met .
\end{enumerate}
A combination of simulation-based, data-based, and simulation-data-mixed prescriptions is used to estimate the background 
rates from the various sources~\cite{single_top}. 
These prescriptions are detailed in Ref.~\cite{WZThesis} and summarized in the following section.

\subsection{Analysis of pretag control region}\label{sec:back_pretag}

The kinematic description of the $W+$jets events, contributed by both $W+$LF and $W+$HF processes, is obtained from
the simulated $W$-plus-$n$-parton samples weighted by their 
LO production cross-sections. As the $W+$jets event yield is predicted with large uncertainties, we use the 
data sample prior to applying $b$-jet-identification requirements  (pretag control region) 
to estimate the total $W+$jets yield and validate the accuracy of the kinematic modeling of
the $W+$jets simulation.

\begin{figure*} [ht]
  \begin{center}
    \subfloat[]{\includegraphics[width=0.49\textwidth]{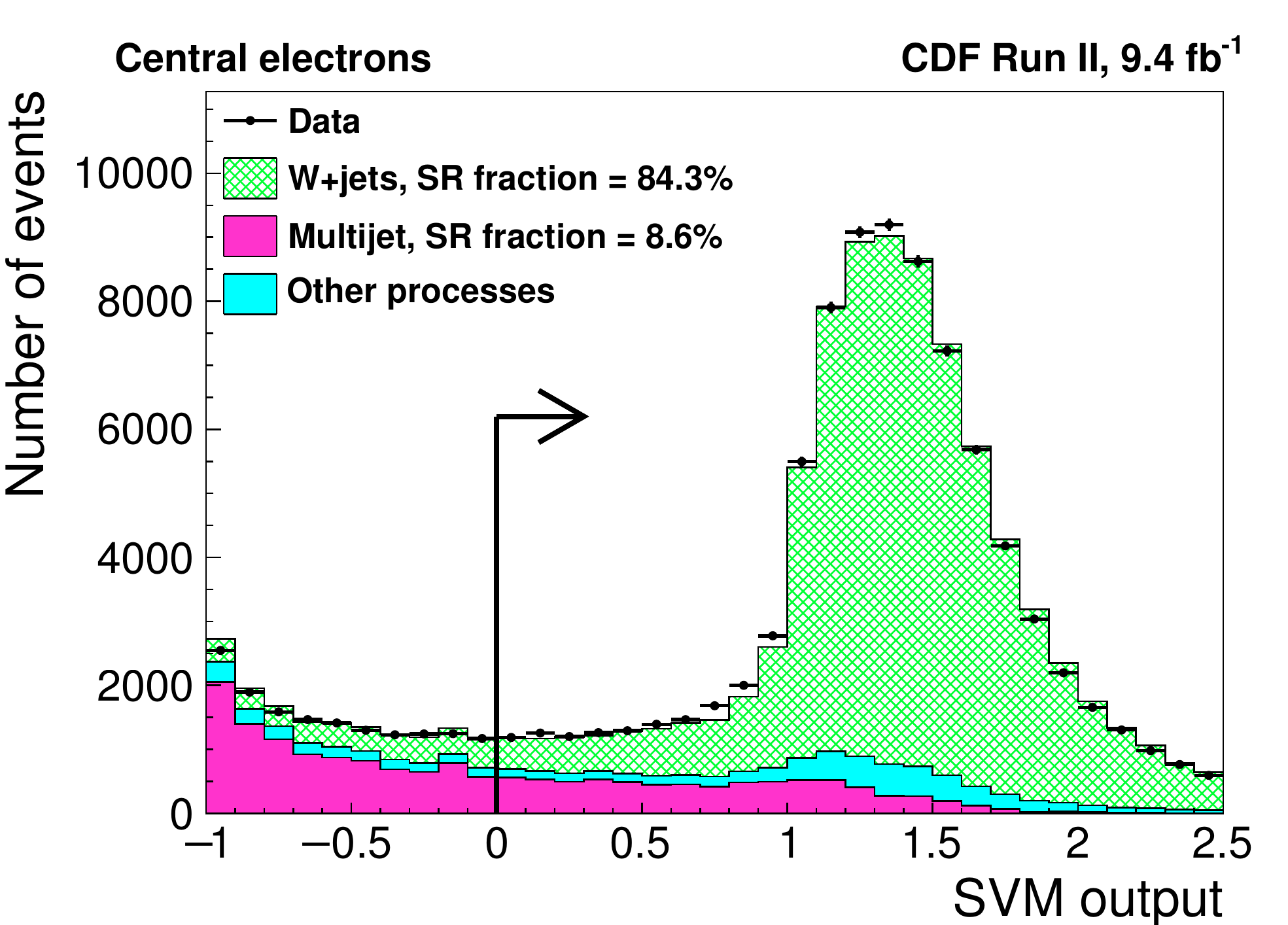}}
    \subfloat[]{\includegraphics[width=0.49\textwidth]{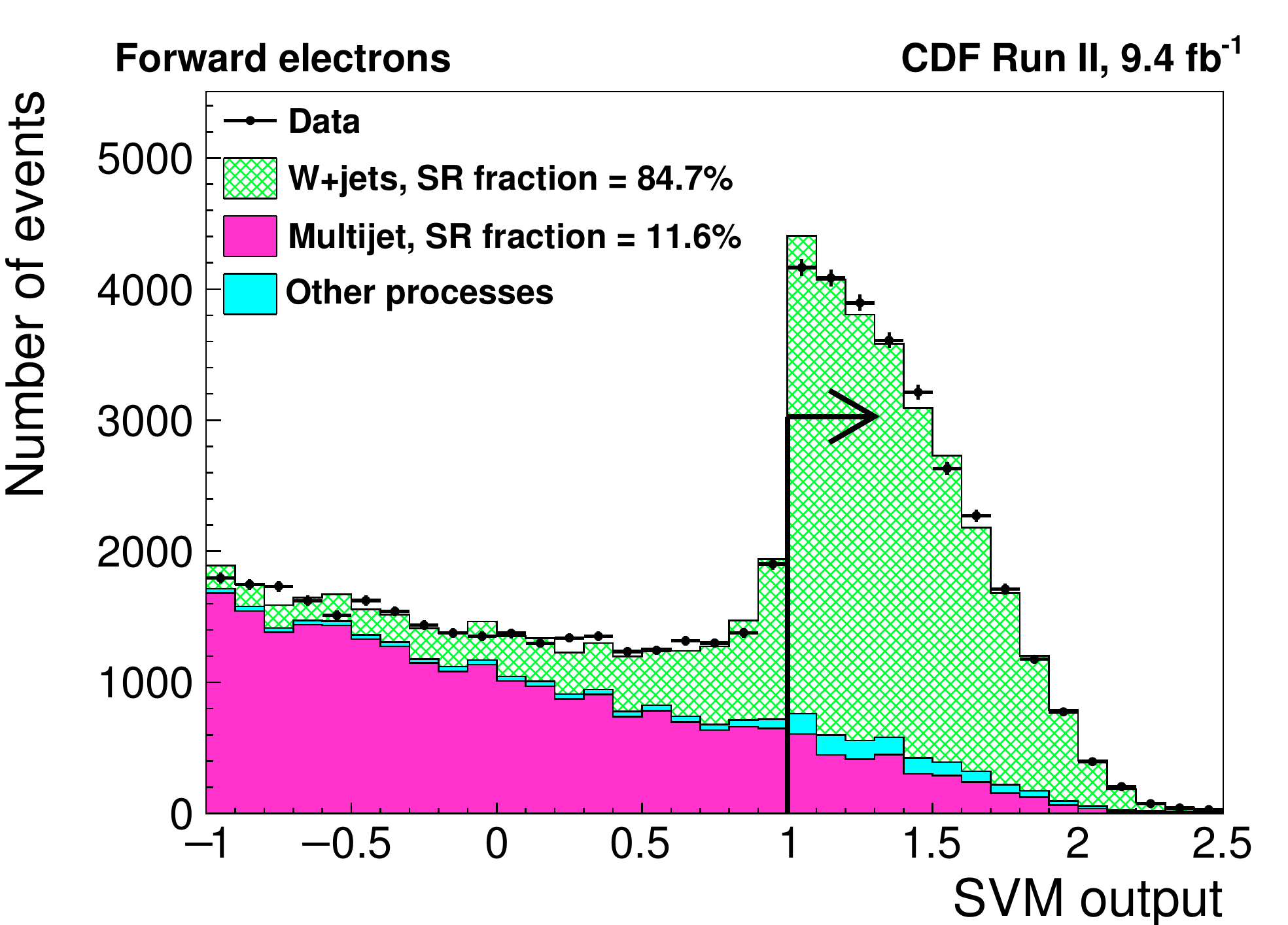}}\\
    \subfloat[]{\includegraphics[width=0.49\textwidth]{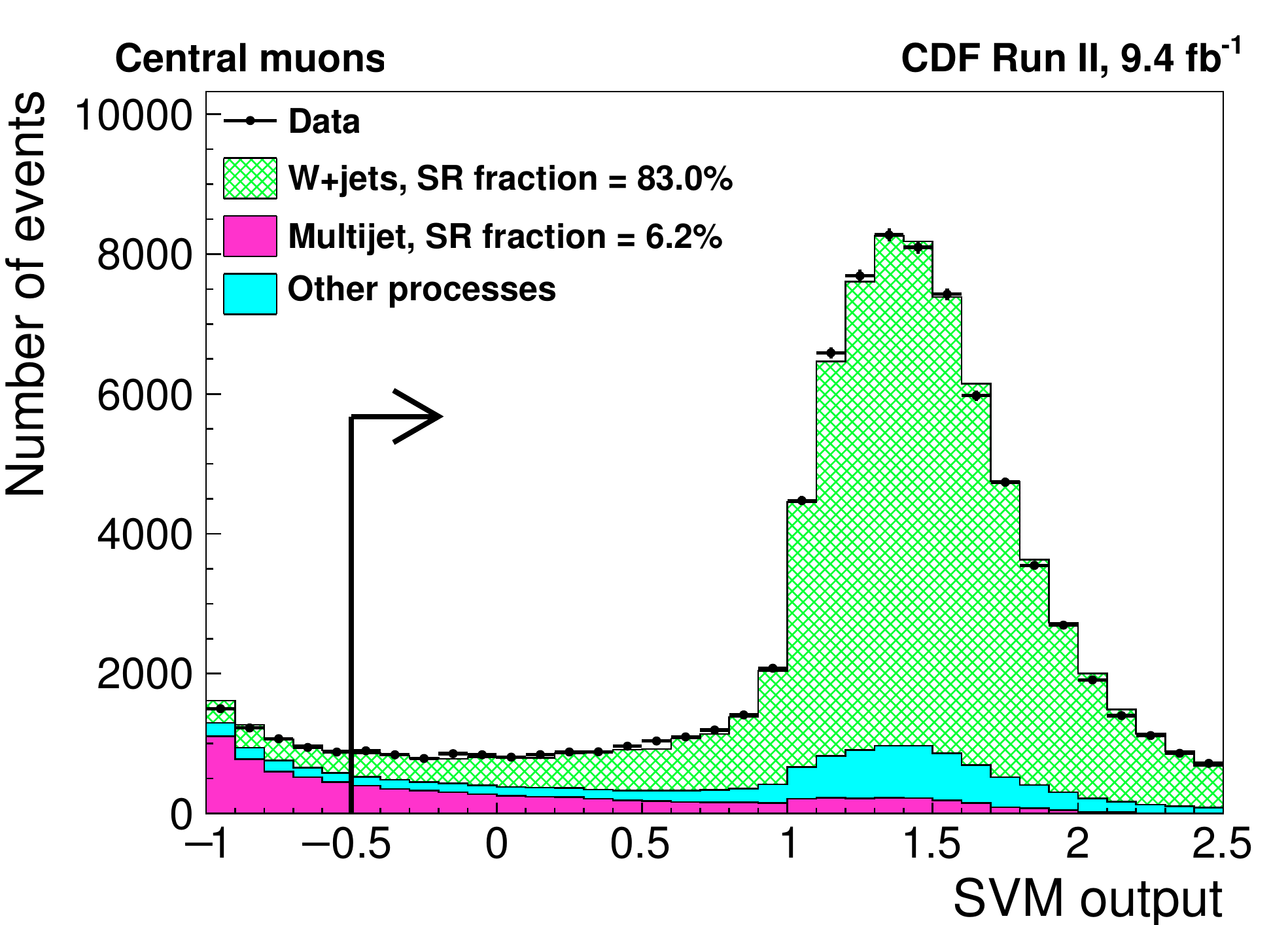}}
    \subfloat[]{\includegraphics[width=0.49\textwidth]{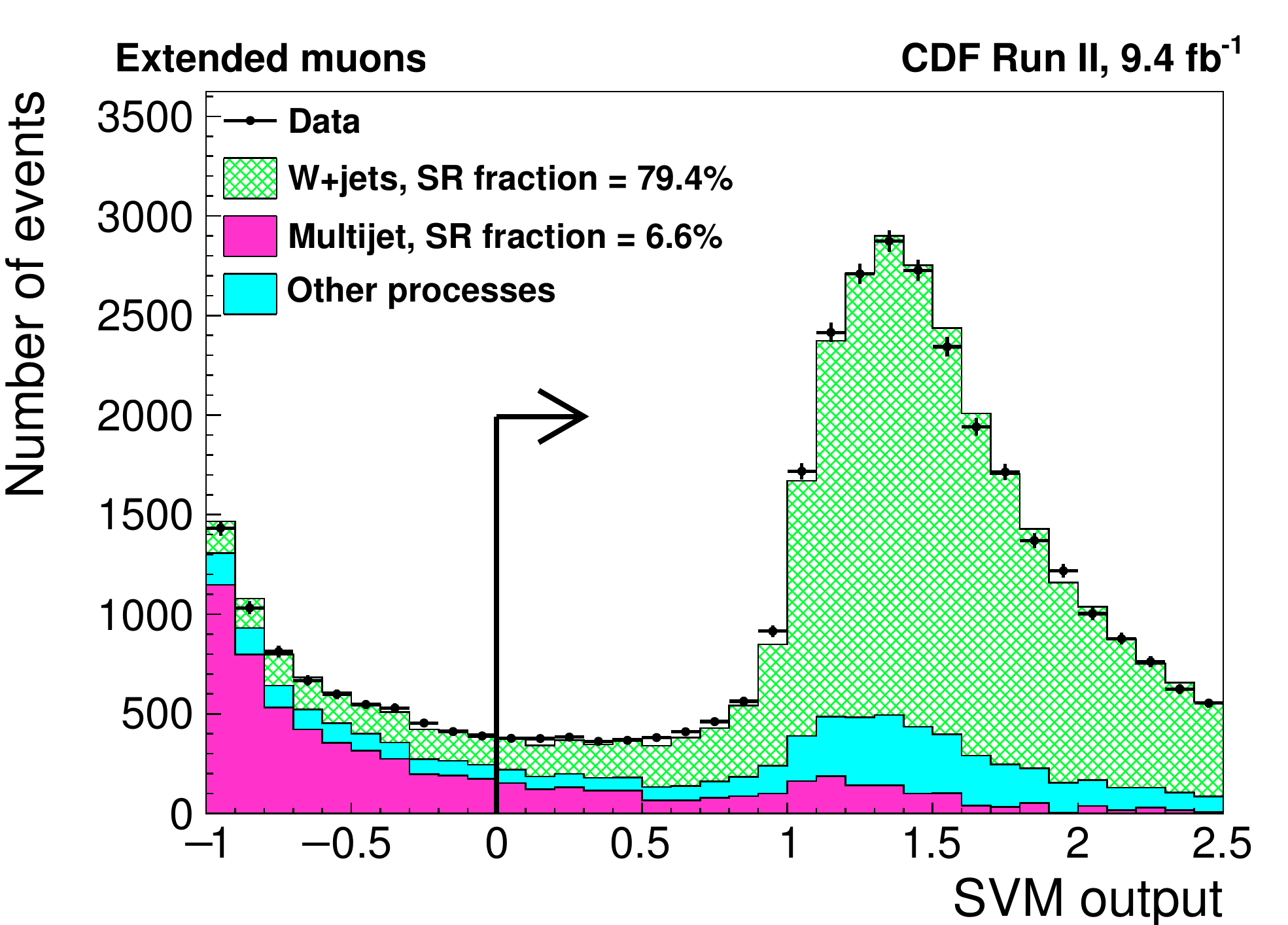}}
\caption{
SVM output distributions for pretag control-region data (points), $W+$jets (light hatched filling), MJ (dark uniform filling), and other processes (light uniform filling) along with the 
MJ and $W+$jets event fractions after the signal region (SR) selection marked by the black arrows. The MJ and $W+$jets event yields 
are determined by fitting backgrounds to the data. The figure shows (a) central electrons, (b) forward electrons, (c) central muons,  and (d) extended muons categories.}  
\label{fig:svm_pretag}
  \end{center}
\end{figure*}

The normalization of the $W+$jets simulation is determined separately in 
each lepton category using a template-likelihood fit of the SVM distribution. Templates 
of the EWK and $W+$jets processes are built using simulation, while the following data-driven models are used for the MJ-background 
templates:
\begin{enumerate}
\item[(1)] Central- and  forward-electrons multijet templates are modeled using a sample obtained by inverting two out of the five selection 
requirements used to define tight electron candidates (see Sec.~\ref{sec:ev_sel}). The \met\, of the central-electron MJ model 
is corrected to account for the 
different calorimeter response of the misidentified electron in data and in the model with inverted selection.
\item[(2)] Central- and extended-muon multijet templates are modeled using a sample obtained by inverting the isolation criteria 
applied to tight- and loose-muon selection algorithms.
\end{enumerate}
The MJ and $W+$jets template normalizations are left free  in the fit, while the EWK components are constrained within their 
uncertainties to normalized predictions using Eq.~(\ref{eq:mc_norm}), together with the theoretical cross sections 
listed in Table~\ref{tab:cx}.
The fit is performed before the application of the SVM selection requirement to leverage the high statistical power 
of the low SVM-output region to constrain the normalization of the MJ background.
Figure~\ref{fig:svm_pretag} shows the results of the fit for the various lepton-class 
 regions together with the different selection thresholds. The fractions of $W+$jets and MJ events 
after the selection requirements are also reported.

\subsection{Background evaluation in the signal regions}\label{sec:HF_bk}

The production of $W$ bosons in association with HF quarks represents the main background process in the single and 
double-tagged signal regions. The kinematic description of the HF-tagged $W+$HF background is derived from simulation 
using \textsc{alpgen}
MC set with massive HF partons used in the matrix-element calculation. 

The $W+$HF normalization is extracted from the simulated HF fractions in the $W+$jets MC after scaling the total $W+$jets MC yield 
 to the one obtained from the pretag control sample. Factors $K_i$ accounts for the $W+$HF yield difference 
in data with respect to the prediction of the HF fractions, 
which results from the interplay of matrix-element generation (at LO), parton-shower matching scheme, 
and the strategy used to avoid HF-parton double-counting across samples.
As described in the Appendix,
the heavy-flavor-fraction correction is derived from a $W+1$ jet control region for 
both $W+b\bar{b}$ plus $W+c\bar{c}$, and  $W+c$ processes. 
In contrast to previous analyses~\cite{single_top,WHbb_2012,signletop_ljet_2014}, the  $W+b\bar{b}$ plus $W+c\bar{c}$, and $W+c$  
corrections are extracted simultaneously. 
The correction factor for $W+b\bar{b}$ plus $W+c\bar{c}$ events is $K_{bb,cc}=1.24\pm 0.25$, and, for $W+c$ events, is $K_c=1.0\pm 0.3$.
The possibility that the corrections $K_i$ are not appropriate for the two-tag selection region, dominated by events with two HF partons identified in well separated jets, is studied at generator level:
an uncertainty of 40\% is used when extrapolating the $K_{bb,cc}$ correction to the two-tag-selection region.

The contamination from $W+$LF events is estimated using the pretag data and a per-jet mistag probability measured in a QCD multijet control 
sample and parametrized as a function of six significant variables (``mistag matrix'').  
The uncertainty in the  $W+$LF event rate is obtained by propagating the systematic uncertainty on 
the per-jet-mistag probability. 
The  kinematic distributions of HF-tagged  $W+$LF events are modeled using the pretag $W+$LF 
component of the simulation and weighting each event for the mistag probability. 

The EWK background predictions are based on simulation with event yields predicted using Eq.~(\ref{eq:mc_norm}) together with the 
theoretical cross sections listed in Table~\ref{tab:cx}.

Finally, the residual MJ component is modeled from the data-driven templates described in Sec.~\ref{sec:back_pretag}, 
plus the additional requirement of selecting events with one or two taggable jets if analyzing the single-tag or double-tag signal regions, 
respectively. The MJ normalization is obtained by a fit to the SVM-output distributions in data using two templates, 
one for the MJ and another for all the other backgrounds. The two template normalizations are free in the fit and the MJ normalization 
result is used in the final background estimate.
A total uncertainty of 40\% is applied to the MJ prediction to account for different MJ data-driven models in the HF-tagged region, 
different boundaries used in the fit on the SVM output, and the use of different variables (e.g., \met) in the template fit.

Table~\ref{tbl:1tag} summarizes the number of observed and expected events in the $W+2$ jets sample, 
for all lepton categories in the pretag, one-tag, and two-tag samples. 

\begin{table*}[!t]
      \begin{center}
    \caption{Summary of  observed and expected event yields in the pretag, one-tag, and two-tag-selection regions, in the $W + 2$ jets sample in 
      9.4 fb$^{-1}$ of integrated luminosity. The single-top process include both $s$ and $t$ channel production modes. The uncertainties include contributions from lepton acceptance, HF-tagging efficiency, luminosity, theoretical uncertainties on EWK backgrounds, mistags estimate, MJ model, and $W+$HF fraction correction.}\label{tbl:1tag}
    \begin{tabular}{cccc}
      \hline\hline
      Process & Pretag    &   one-tag            & two-tag \\
      \hline
MJ               & $18\,100$ $\pm$	2700  & 800	$\pm$	330  &  30	$\pm$	14 \\ 
$W+$LF           & $161\,700$ $\pm$	3700  & 2440	$\pm$	350  &  29.5	$\pm$	6.8 \\ 
$W+c\bar{c}$     & $13\,400$ $\pm$	1700  & 1190	$\pm$	290  &  33	$\pm$	16 \\ 
$W+c$            & $11\,600$ $\pm$	2200  & 930	$\pm$	310  &  12.5	$\pm$	5.5 \\ 
$W+b\bar{b}$     &6370  $\pm$	930   & 2190	$\pm$	520  &  313	$\pm$	125 \\ 
$Z+\textrm{jets}$           &9400  $\pm$	1900  & 281	$\pm$	42   &  13.5	$\pm$	2.1 \\  
$t\bar{t}$       &1600  $\pm$	230   & 663	$\pm$	94   &  137	$\pm$	22 \\ 
Single-top &1109  $\pm$	42    & 441	$\pm$	23   &  70.8	$\pm$	8.4 \\ 
{\it ZZ}             &93.4	$\pm$	4.4   & 10.1	$\pm$	0.7    &  2.0	$\pm$	0.3  \\ 
{\it WH+ZH}         &40.0	$\pm$	1.4   & 17.6	$\pm$	0.8    &  5.4	$\pm$	0.6 \\ 
{\it WW}             &5530  $\pm$	400   & 240	$\pm$	30   &  3.0	$\pm$	0.7 \\ 
{\it WZ}             &904	$\pm$	53    & 91.4	$\pm$	7.6    &  17.2	$\pm$	2.1 \\ 
      \hline
Total prediction  & $229\,900\pm 5800$ & $9300 \pm 1200$ &  $670 \pm 140$\\
Observed data    & 232\,145 & 9074	               &  604	\\                        
\hline\hline
    \end{tabular}      
  \end{center}
\end{table*}

\section{{\it WW} and {\it WZ} signal discrimination}\label{sec:discriminant}

After the HF-tag requirement the expected signal-to-background ratio is less than 0.04. 

However, additional sensitivity comes from the study of 
 the distribution of the invariant mass of the two jets in the event, $m_{jj}$, where
 signal is expected to cluster in a narrow resonance structure over a smooth nonresonant background.
To improve the poor invariant mass resolution  of the hadronic final state, the jet energy is corrected with a 
neural-network-based calibration~\cite{timoNN}, which uses information from jet-related variables and from the secondary decay vertex, 
if reconstructed in a jet. The dijet-invariant-mass resolution, initially about 15\%, is improved to about 13\% and 11\% in the single 
and double-tag signal regions, respectively.

\begin{figure*} [!ht]
\centering
\subfloat[]{\includegraphics[width=0.49\textwidth]{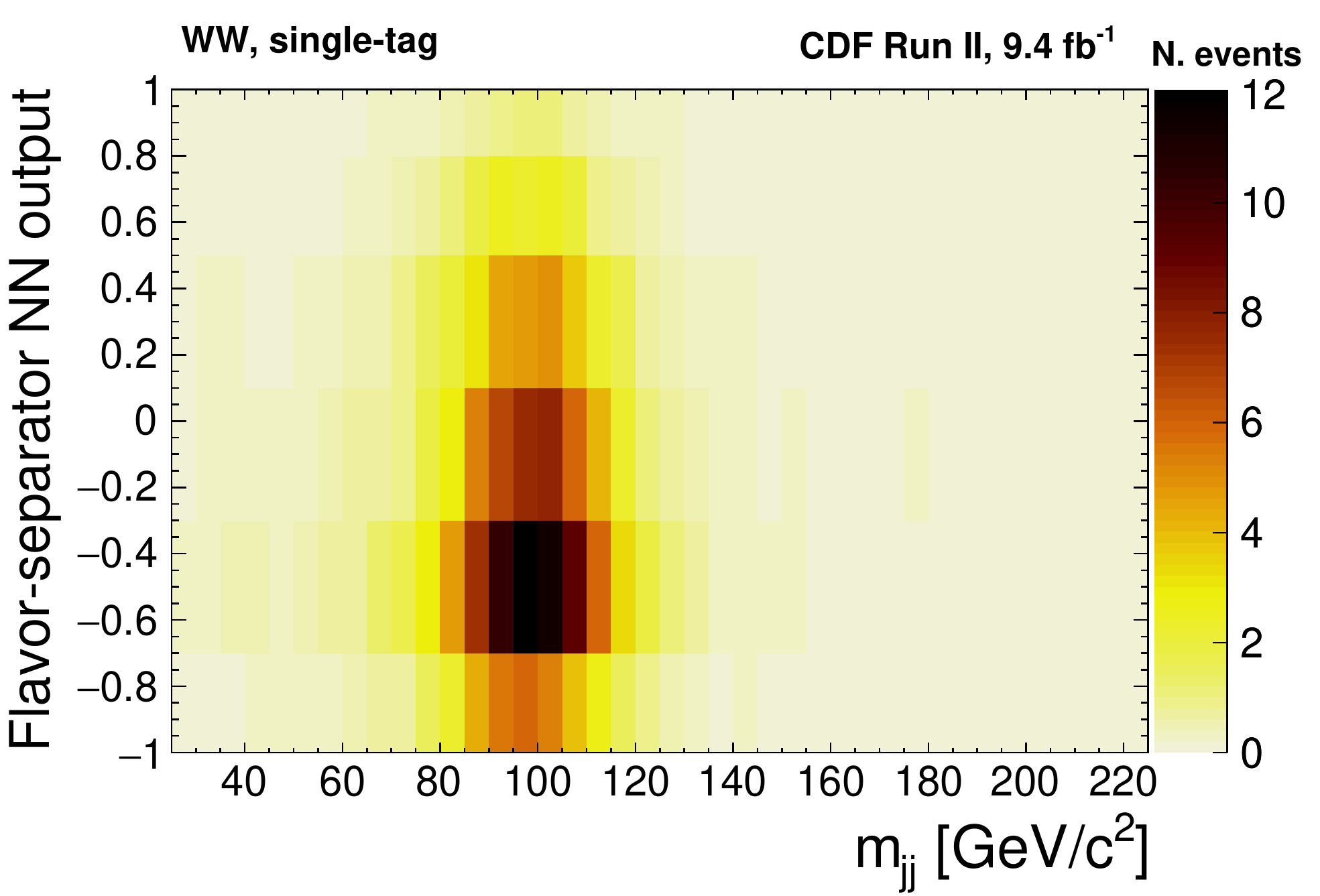}}
\subfloat[]{\includegraphics[width=0.49\textwidth]{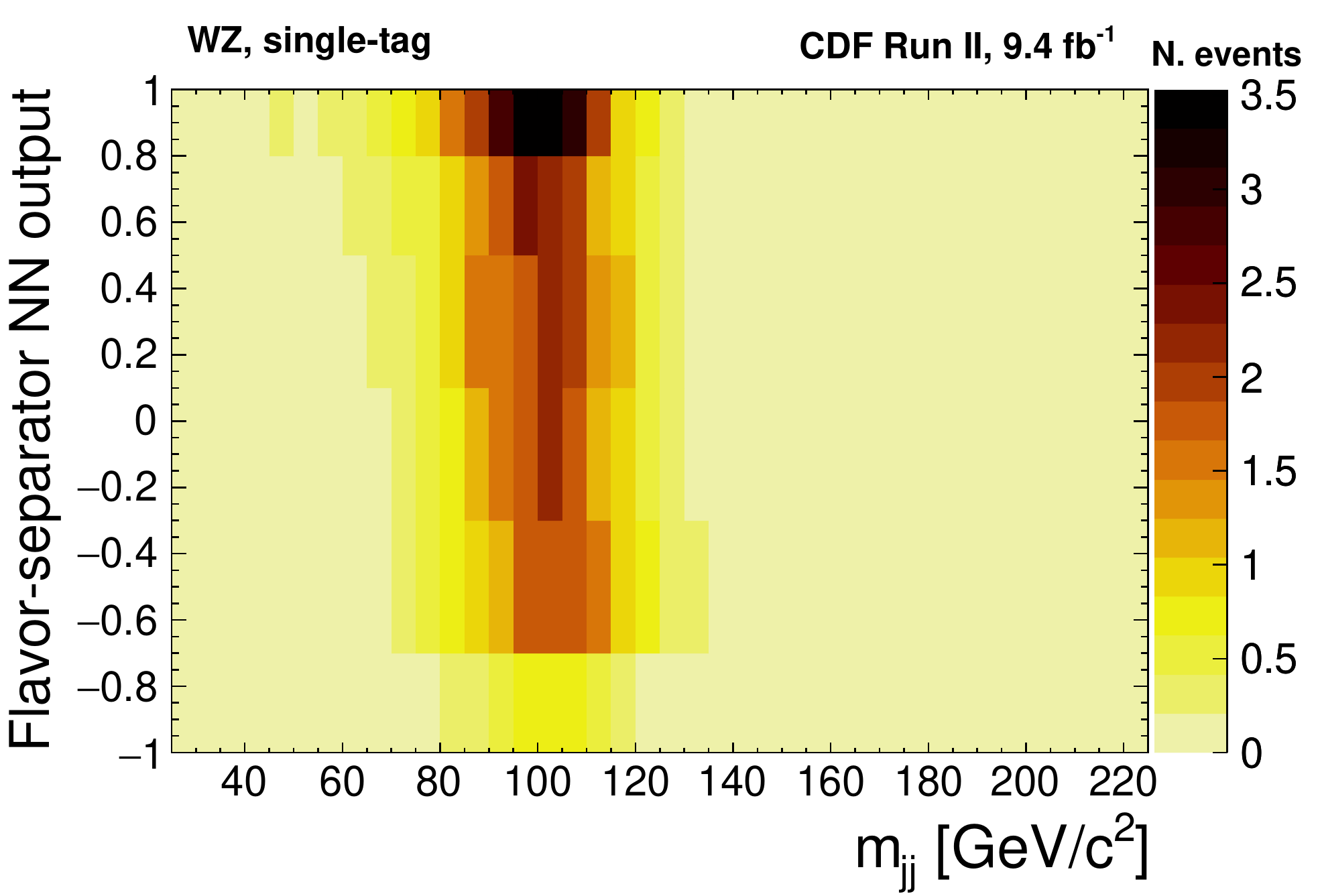}}\\
\subfloat[]{\includegraphics[width=0.49\textwidth]{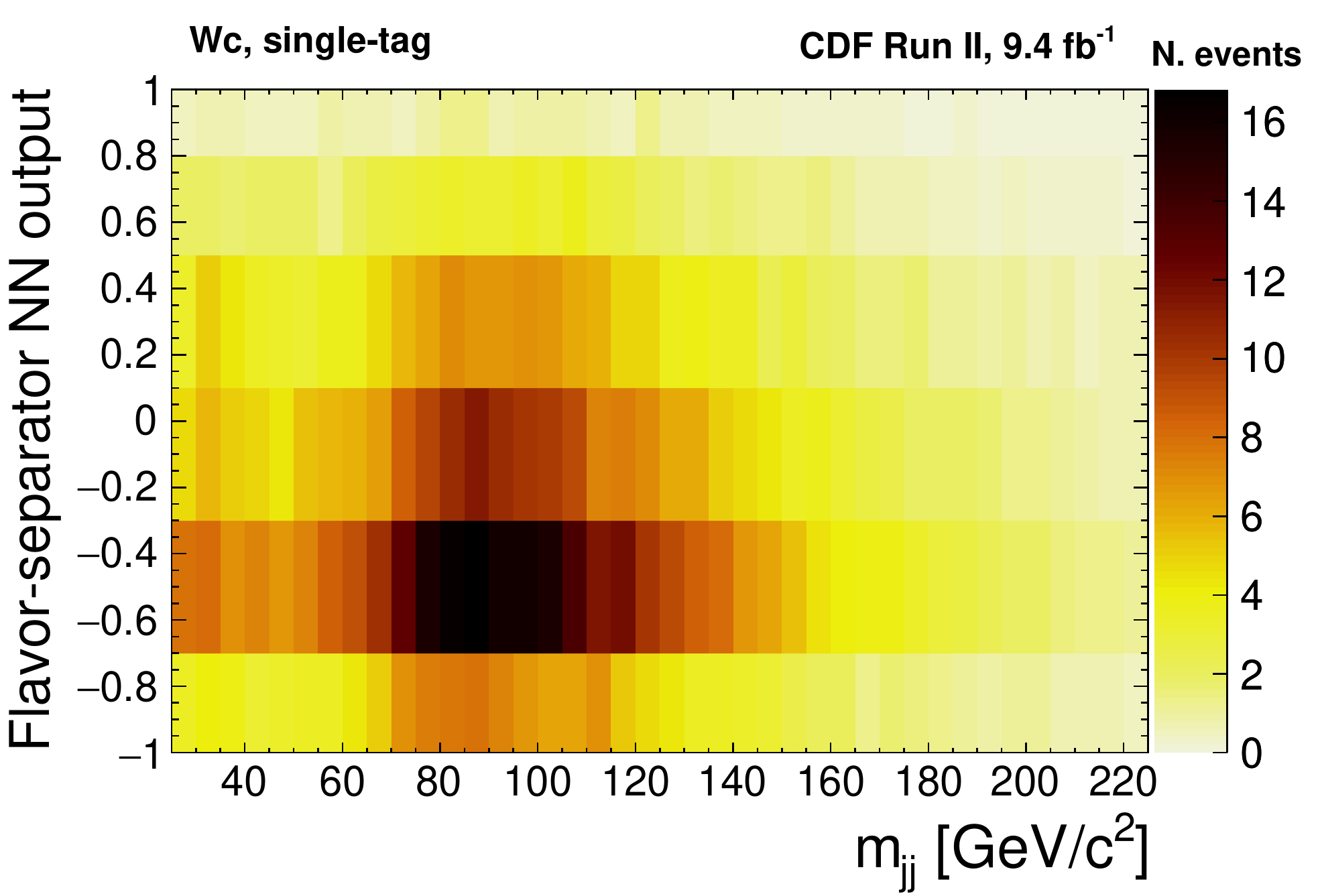}}
\subfloat[]{\includegraphics[width=0.49\textwidth]{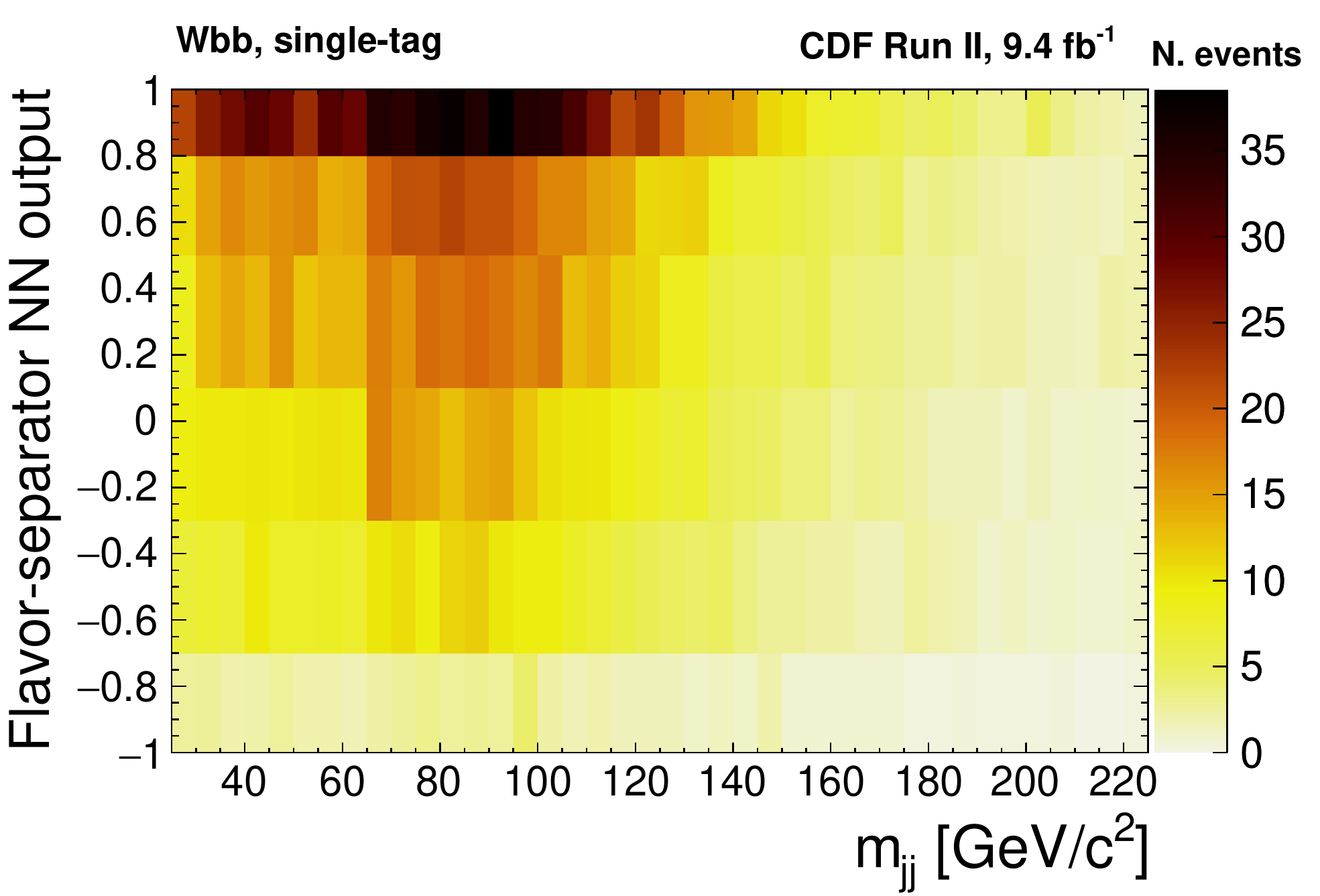}}
\caption{Examples of two-dimensional templates of $m_{jj}$ versus flavor-separator-NN output for (a) {\it WW} signal, (b) {\it WZ} signal, (c) $W+c$ background, and (d) $W+b\bar{b}$ background in single-tagged $W+2$ jets events.}  
\label{fig:kit_mjj}
\end{figure*}

For single-tag events, in addition to the signal-to-background discrimination power of the $m_{jj}$ distribution, the flavor-separator 
NN~\cite{single_top}  is used to achieve $b$-to-$c$-jet separation. 
The flavor-separator NN uses the information from the secondary-decay vertex to assign an output 
score between --1 and 1, depending 
on the jet being more 
$b$-like or LF-like. Jets originating from $c$ quarks are likely to obtain negative scores, clustering around the NN output value of $- 0.5$.  
The $m_{jj}$ distribution and the flavor-separator NN output (divided in six bins) are combined in a two-dimensional distribution. 
This improves the separation of the {\it WW} and the {\it WZ} signals in the single-tag signal regions.
Figure~\ref{fig:kit_mjj} shows example distributions for  signals and  background processes.

The discrimination power of the flavor-separator NN is not important in the two-tag signal region as 
 approximately 90\% of the events selected in this category are expected to originate from processes with $b$ quarks in the final state. 
Therefore only the $m_{jj}$ distribution is used for the signal extraction in the two-tag signal region.

\begin{figure*} [!ht]
\centering
\subfloat[]{\includegraphics[width=0.49\textwidth]{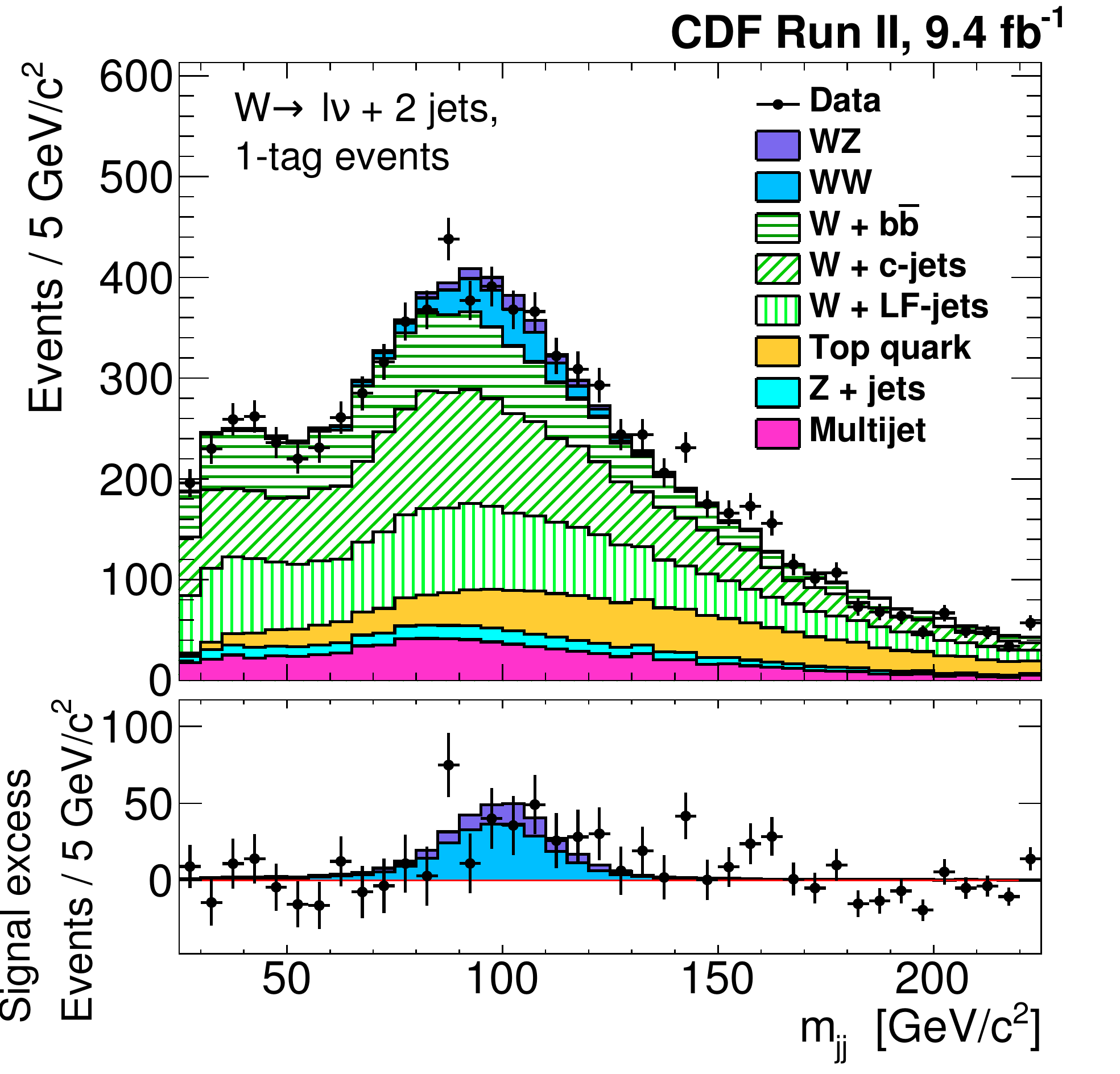}}
\subfloat[]{\includegraphics[width=0.49\textwidth]{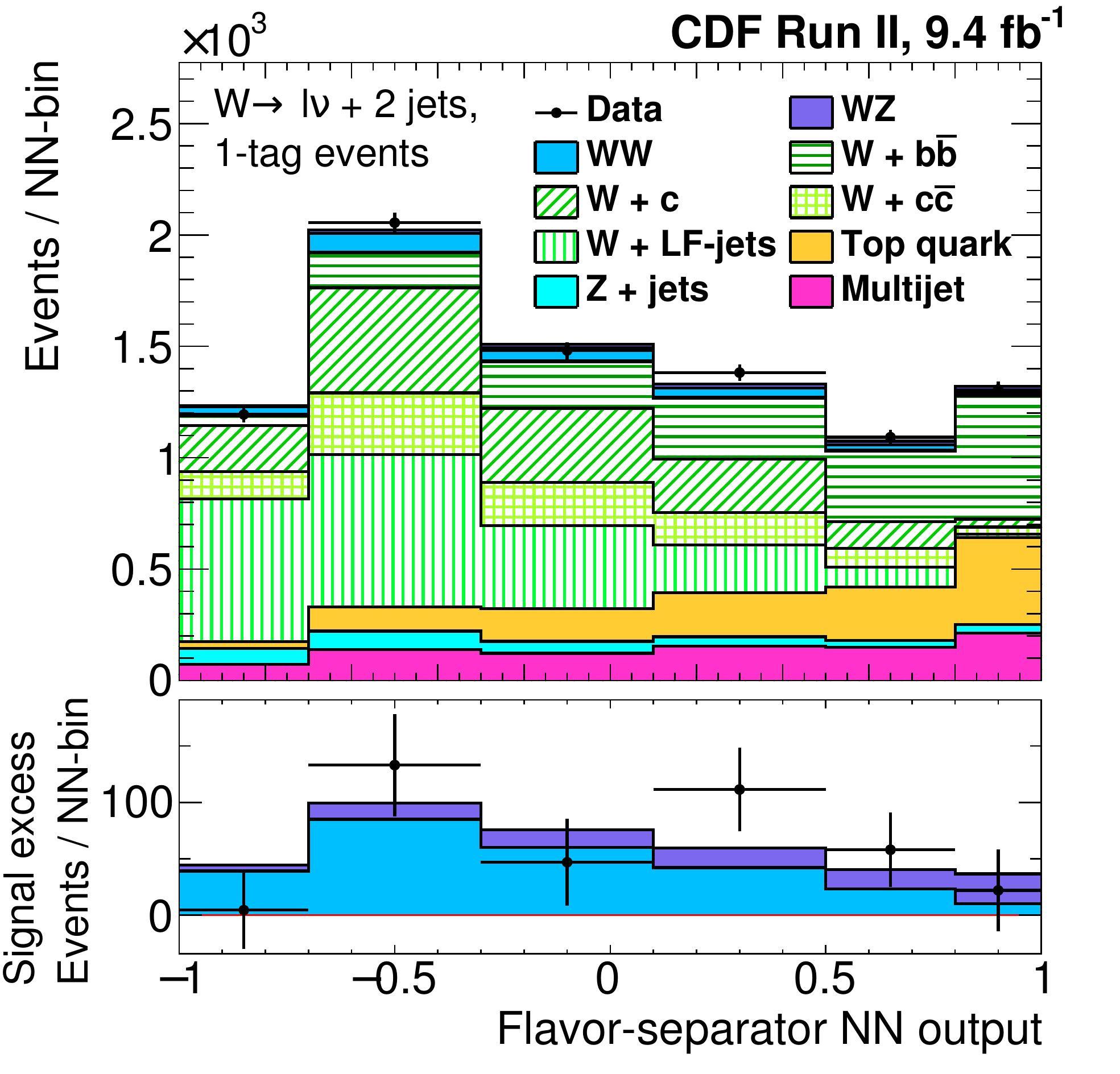}}
\caption{Distributions of  $m_{jj}$ and flavor-separator-NN output for the one-tag candidates where events from all
 lepton categories are added together.
Rate and shape systematic uncertainties of signal and background processes are treated as nuisance parameters and the
 best fit to the data is shown.}
\label{fig:mjj_1tag}
\end{figure*}

\begin{figure*} [!ht]
\centering
\subfloat[]{\includegraphics[width=0.49\textwidth]{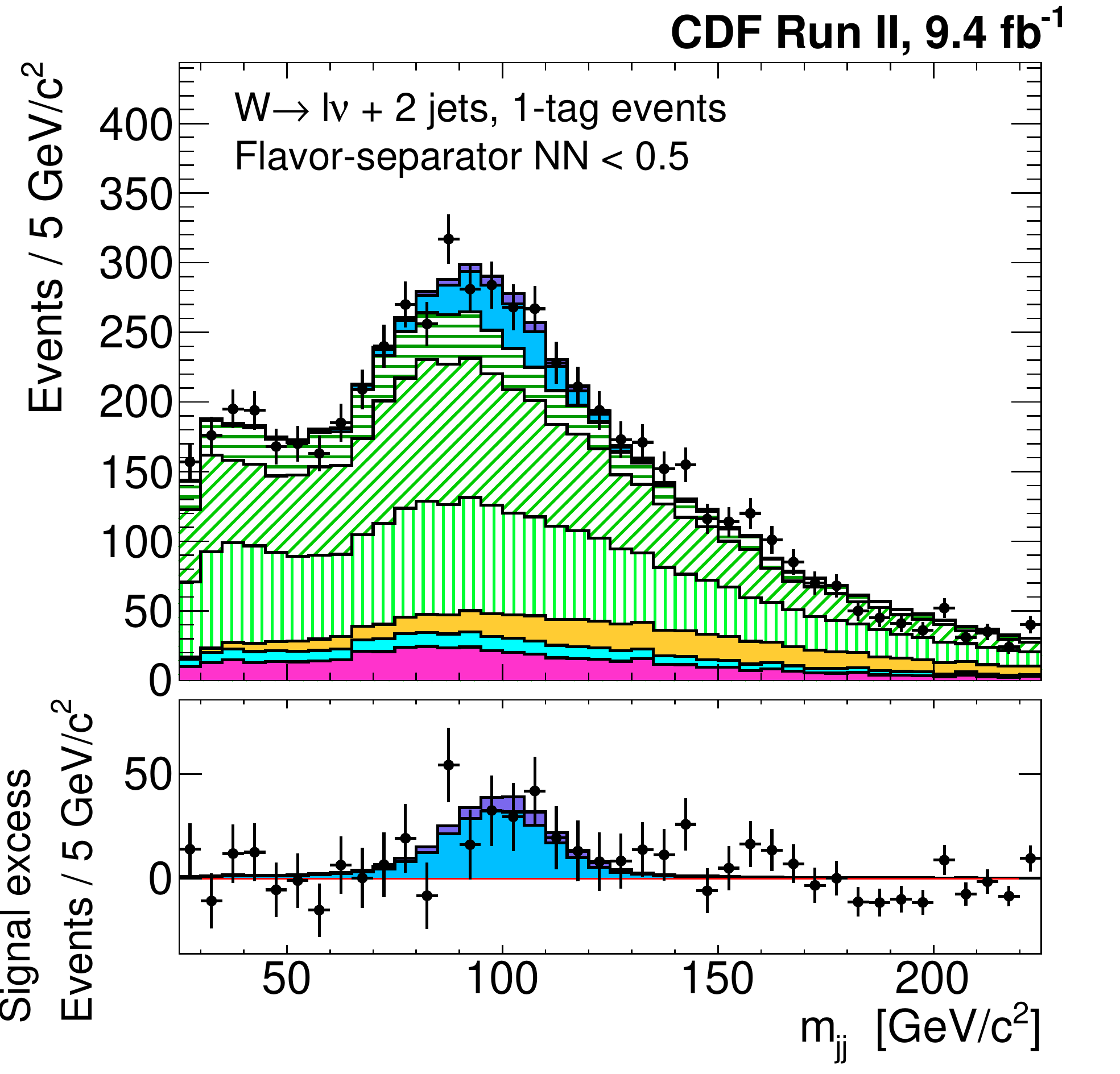}}
\subfloat[]{\includegraphics[width=0.49\textwidth]{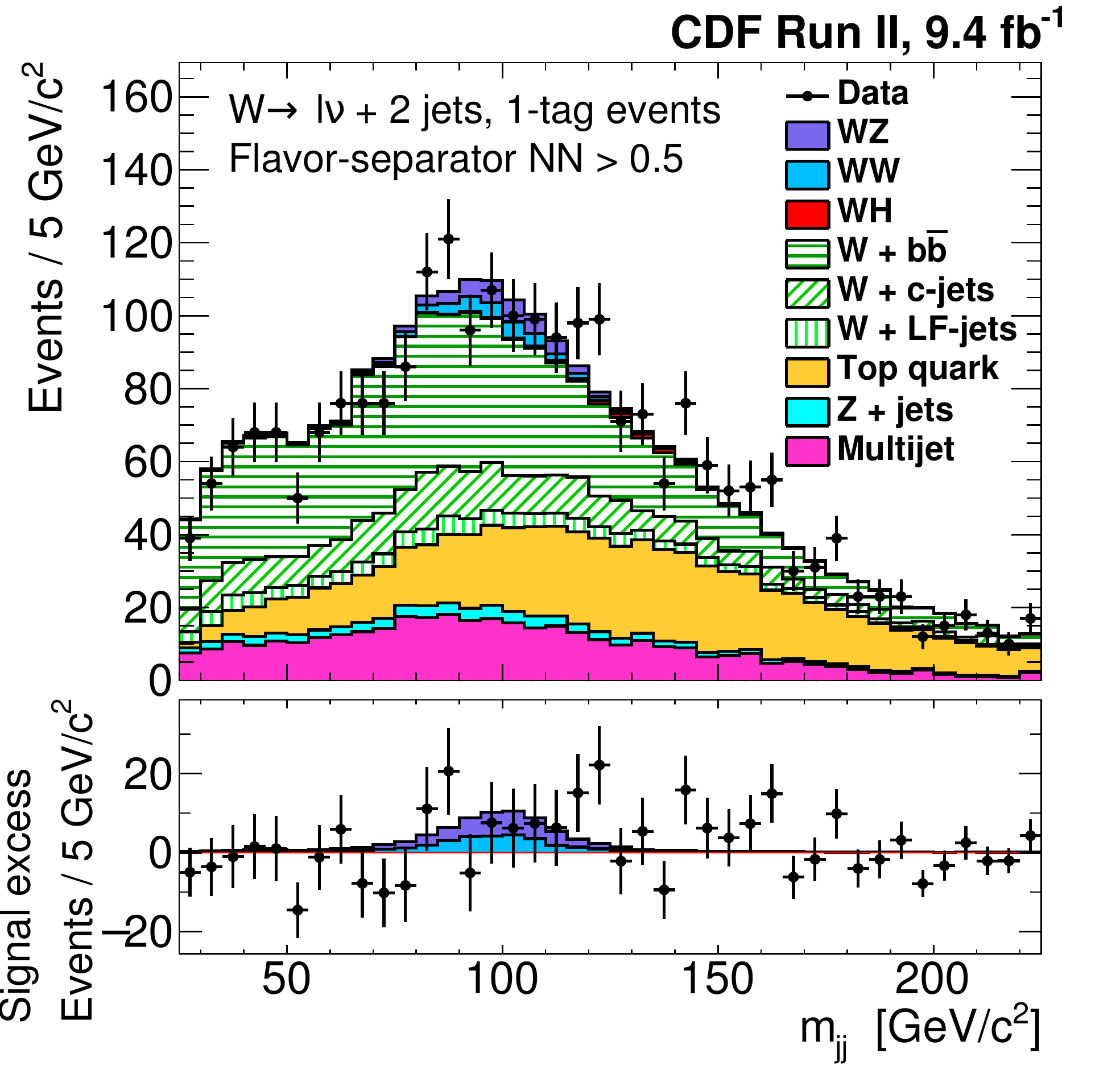}}
\caption{Distributions of $m_{jj}$  for the one-tag candidates;
for events with (a) flavor-separator-NN output$<0.5$, and (b) flavor-separator-NN output$>0.5$. Events from all  lepton categories 
are added together. Rate and shape systematic uncertainties of signal and background processes are treated as nuisance parameters 
and the best fit to the data is shown.}
\label{fig:mjj_1tag_split}
\end{figure*}

\begin{figure} [!ht]
\centering
\includegraphics[width=0.49\textwidth]{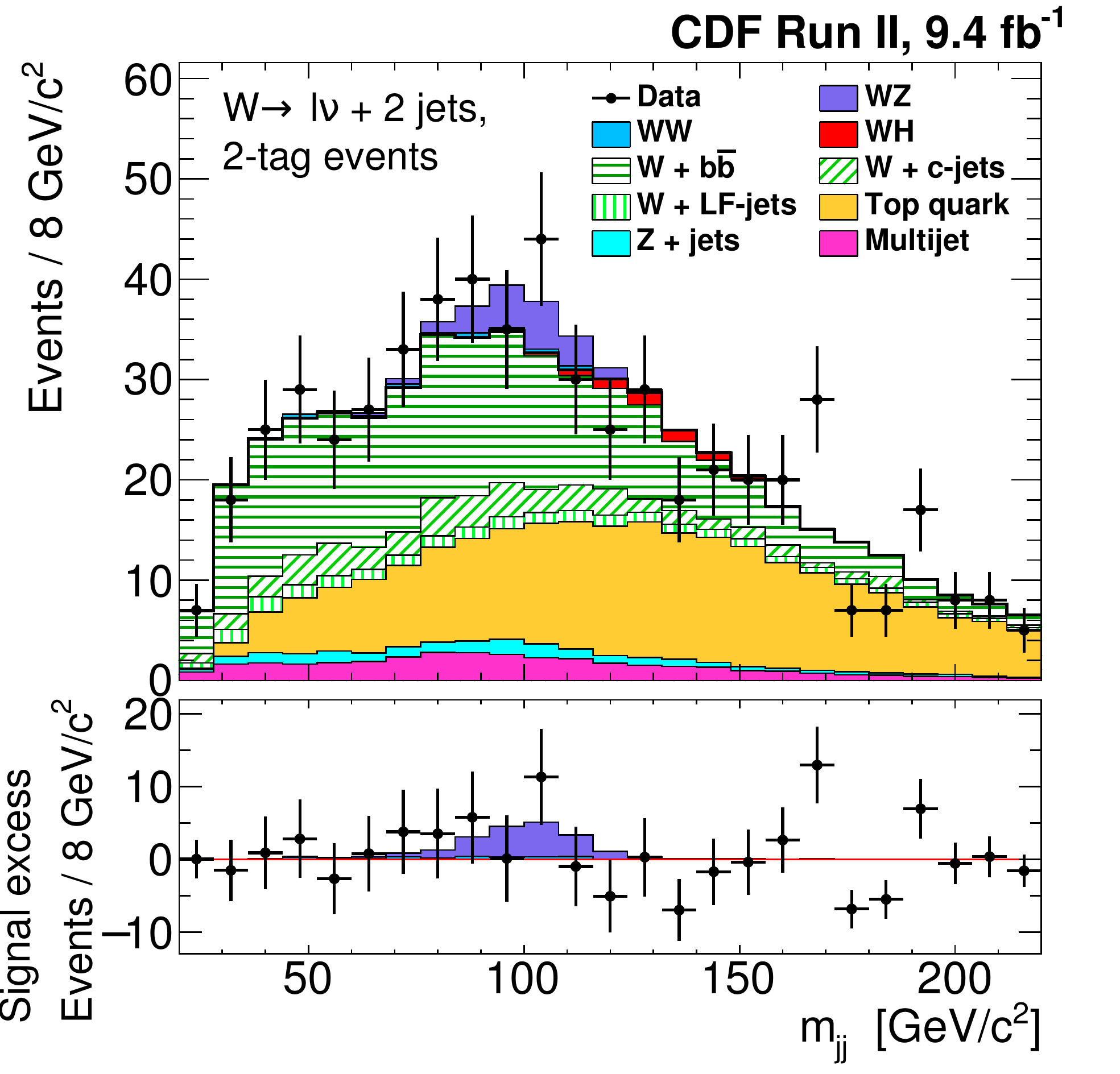}
\caption{Distributions of  $m_{jj}$  for the two-tag candidates where events from all  lepton categories are added together.
Rate and shape systematic uncertainties of signal and background processes are treated as nuisance parameters and 
the best fit to the data is shown.}
\label{fig:mjj_2tag}
\end{figure}

In total, eight regions are used for the signal extraction: four lepton subsamples (central electrons, central muons, forward 
electrons, extended muons) times two HF-tag prescriptions (one tag with flavor-separator NN and two tags). 
The distributions for all the lepton categories restricted to the one-tag final state are shown in Fig.~\ref{fig:mjj_1tag}: 
In panel (a) they are integrated across the flavor-separator-NN output and projected onto $m_{jj}$, while in panel (b) the 
integration 
is across $m_{jj}$ and the projection is onto the flavor-separator-NN output distribution. Figure~\ref{fig:mjj_1tag_split} 
shows the single-tag-$m_{jj}$ distribution for the $b$-enriched-flavor-separator-NN region 
(flavor-separator-NN output $>0.5$, Fig.~\ref{fig:mjj_1tag_split}a) and $b$-suppressed region (flavor-separator-NN output $<0.5$, Fig.~\ref{fig:mjj_1tag_split}b).
Finally, the $m_{jj}$ distribution for two-tag events, summed for all the lepton categories,
is shown in  Fig.~\ref{fig:mjj_2tag}.

\section{Cross section measurements and statistical analysis}\label{sec:fit}

In order to measure the total ({\it WW+WZ} ) and separate {\it WW} and {\it WZ} production cross-sections in the HF-enriched final state, 
data distributions described in Sec.~\ref{sec:discriminant} are compared to expectations of signal and background using a Bayesian analysis, 
following Ref.~\cite{single_top}.

A likelihood function is built from the
observed numbers of events in each bin, and the estimated signal and
background distributions, assuming Poisson statistics. 
The prior probabilities of background and signal templates are included in the likelihood, together with all rate and 
shape systematic uncertainties, which are treated as nuisance parameters.  The unknown yields of 
the two signal processes 
({\it WW} and {\it WZ} ) are parametrized by  uniform prior distributions in the non-negative domain. 
The signal production total cross-section is obtained 
by marginalizing the prior probability distribution over the nuisance parameters and
studying the resulting posterior distribution for the signal yield. As the  
signal parametrization is normalized to the SM expectation, the maximum value of the Bayesian posterior corresponds to the 
measurement of the signal strength $\mu$,
\begin{equation}
  \label{eq:sgn_strength}
  \mu^{\textrm{obs}} = (\sigma \times {\cal B})^{\textrm{obs}}_{\textrm{signal}}/ (\sigma \times {\cal B})^{\textrm{SM}}_{\textrm{signal}}.
\end{equation}
Half of the shortest interval enclosing 68.3\% (95.5\%) of the posterior integral provides 
$1\sigma$ ($2\sigma$) uncertainty at the corresponding Bayesian credibility level (C.L.).
For the combined {\it WW+WZ} result, a one-dimensional uniform signal prior is used with relative 
rates for the {\it WW} and {\it WZ} processes given by the SM.
When measuring the {\it WW} and {\it WZ} cross sections separately, a two-dimensional uniform prior is used.

\subsection{Systematic uncertainties}\label{sec:sys}

Systematic uncertainties,
which may affect both the yield (rate uncertainties) and the distribution of the discriminant variables (shape uncertainties),
are used to account for the limited detector resolution and accuracy of calibrations,
extrapolations from control regions, and theoretical predictions.
Rate uncertainties are included in the likelihood by assigning a Gaussian prior probability to the normalization of a given process. 
If a given uncertainty source has relevant impact on the $m_{jj}$ or flavor-separator-NN output shapes, 
these distributions are modified and the relative difference between the varied and the nominal 
distributions are included in the likelihood as correlated bin-by-bin Gaussian variations~\cite{single_top}. 
Statistical fluctuations may be large when evaluating the difference of two distributions, so the bin-by-bin variations affecting 
the $m_{jj}$ distribution are smoothed with a three-bin median filter algorithm~\cite{Filtering}. 
This smoothing choice preserves shape correlations among the distributions if their effect extends through a few consecutive bins.

The systematic uncertainties affecting the rate of signal or background processes 
and considered for this analysis are the following:
\begin{enumerate}
\item[(1)]{\em Luminosity.} A 6\% uncertainty is applied to the expected rate of signal and EWK backgrounds 
based on the 4.0\% uncertainty in the extrapolation of the inelastic
$p\bar{p}$ cross section~\cite{ppbar_cx} and the 4.4\% uncertainty in the acceptance
of the luminosity monitor~\cite{clc_performance}.
\item[(2)]{\em  Lepton acceptance.} Such sources comprise the uncertainties arising from the measurement of the trigger efficiencies, and of the lepton-reconstruction scale-factors.
The expected rate of signal and EWK backgrounds is affected by an uncertainty ranging from 2\% (central muon final state) to 5\% (extended muon final state), 
uncorrelated across the four lepton final states.
\item[(3)]{\em  $b$- and $c$-tagging efficiences.} The uncertainty arising from the per-jet HF-tagging efficiency scale-factor, 
uncorrelated between $b$- and $c$-quark jets, 
is propagated to the final yield of each process.
Rate variations for processes with $b$ quarks in the final state range from approximately 3\% (in cases of events with two $b$ jets selected in the single-tag samples) 
to approximately 10\% (in cases with two $b$ jets selected in the double-tag samples).  About twice such uncertainties are applied to processes with  
$c$ quarks in the final state~\cite{single_top}.
This results in approximately a $10$\% rate uncertainty for processes  selected in the single-tag signal region with one $c$ quark in the final state.
\item[(4)]{\em  PDFs and radiative corrections.} Uncertainties on the signal acceptance due to the choice of 
PDF is evaluated following Ref.~\cite{WH7.5}. 
  Uncertainties on the signal simulation due to the initial- (ISR) and final-state radiation (FSR) are evaluated by halving and doubling the ISR and FSR parameters and considering the resulting signal-acceptance and shape variations. 
The resulting shape variations are small and are neglected. The signal-rate variations are added in quadrature 
for a total systematic uncertainty of approximately 4\%.
\item[(5)]{\em  Theory uncertainties on EWK backgrounds.} Theoretical uncertainties on production cross-sections and acceptances 
of the EWK background processes have  
rate uncertainties of 10\% for top-quark-related processes, 5\% for {\it WH} and {\it ZH} production, 6\% for {\it ZZ} production, 
and 40\% for $Z$+jet. 
The large uncertainty in the $Z+$jets production is due to the conservative uncertainty assigned to the $Z+$HF rate predictions of the \textsc{alpgen} simulation.
\item[(6)]{\em  Mistag estimate.} The uncertainties associated with the mistag matrix are propagated to the $W+$LF yield predictions in the signal regions. 
The resulting rate variations are 15\% and 23\% for the single- and double-tag samples, respectively, and are correlated.
\item[(7)]{\em  $W+$HF fractions corrections.} $W+b\bar{b}$ plus $W+c\bar{c}$, and $W+c$ yields are varied according to the uncertainties associated with 
the $K-$factor corrections described in Section~\ref{sec:HF_bk}. Uncorrelated rate variations of 30\%, 20\%, and 40\% are applied to the 
predictions 
of  $W+c$ in the single-tag sample, $W+b\bar{b}$ plus $W+c\bar{c}$  in the single-tag sample, and $W+b\bar{b}$ plus $W+c\bar{c}$ in the 
double-tag sample, respectively.
\end{enumerate}
Additional systematic uncertainties affecting both the rate and the shape of signal or background processes 
and considered for this analysis are the following:
\begin{enumerate}
\item[(8)]{\em  Flavor-separator NN response to $c$- and LF-quark jets.} 
Decay vertices originating from $c$ and LF hadrons typically have lower track multiplicity and 
smaller distance from the primary vertex than the ones 
originating from $b$ hadrons. As a consequence they are more sensitive to track resolution and detector-material effects, which are difficult to simulate precisely.
Therefore, following the prescriptions of Ref.~\cite{single_top},  an additional uncertainty is 
assigned to the flavor-separator-NN output  distribution of the processes containing  $c$ and LF 
quarks:
Their heavy flavor-separator-NN output shapes are varied between the ones obtained from the 
simulation and the shapes observed in a multijet-data sample enriched in misidentified HF tags.
\item[(9)]{\em  MJ model.} A rate uncertainty of 40\% is applied to each MJ estimate in the various lepton final-states. The shape of the flavor-separator-NN output distribution of the MJ is extracted from templates of different quark flavor following Ref.~\cite{single_top}.
\item[(10)]{\em  Jet energy-scale corrections.} The $1\sigma$ envelope of the uncertainties arising from the 
jet energy-scale corrections is evaluated for each jet as described in Ref.~\cite{jet_corr1}. 
The change in rate and shape of every process is taken into account and new iterations of the complete background estimate 
are performed using the varied samples. 
The resulting rate systematic uncertainty ranges from a few percent to approximately $30$\% depending on the process and on the final state. 
The shape variation produces a shift of up to 3\% in the $m_{jj}$ signal position.
\item[(11)]{\em $W+$jets  $Q^2$.}   \textsc{alpgen} is a LO generator and its  modeling is heavily affected by the choice of the factorization and 
renormalization scale, $Q^2$.
The nominal scale used for $W+n$ parton processes is given by $Q^2 = M^2_W + \sum^{j}p^2_T$,
where $M_W$ is the $W$-boson mass, the sum runs over the partons, and $p_T$ is the transverse energy of each parton.
Various simulated samples of $W+$jets, with $Q$ halved or doubled, are used to account for the scale-choice uncertainty.
Such samples have different kinematic properties and flavor composition. 
These are used for a new iteration of the complete background estimate and the 
difference in rate and shape with respect to the nominal estimate is used for the uncertainties of the $W+$LF and $W+$HF processes. 
\end{enumerate}

After the marginalization, the largest systematic effects on the {\it WW+WZ} cross-section measurement 
are given by the uncertainties on the HF-tagging efficiencies and on the
 $W+$HF fraction corrections, each contributing about 10\% to the uncertainty of the measurement. 

\subsection{Results}
The {\it WW+WZ} cross section is first measured 
by constraining the relative fraction of the {\it WW} and {\it WZ} components to the SM values and by studying the 
total yield of the two processes.

\begin{figure} [!ht]
\centering
\includegraphics[width=0.49\textwidth]{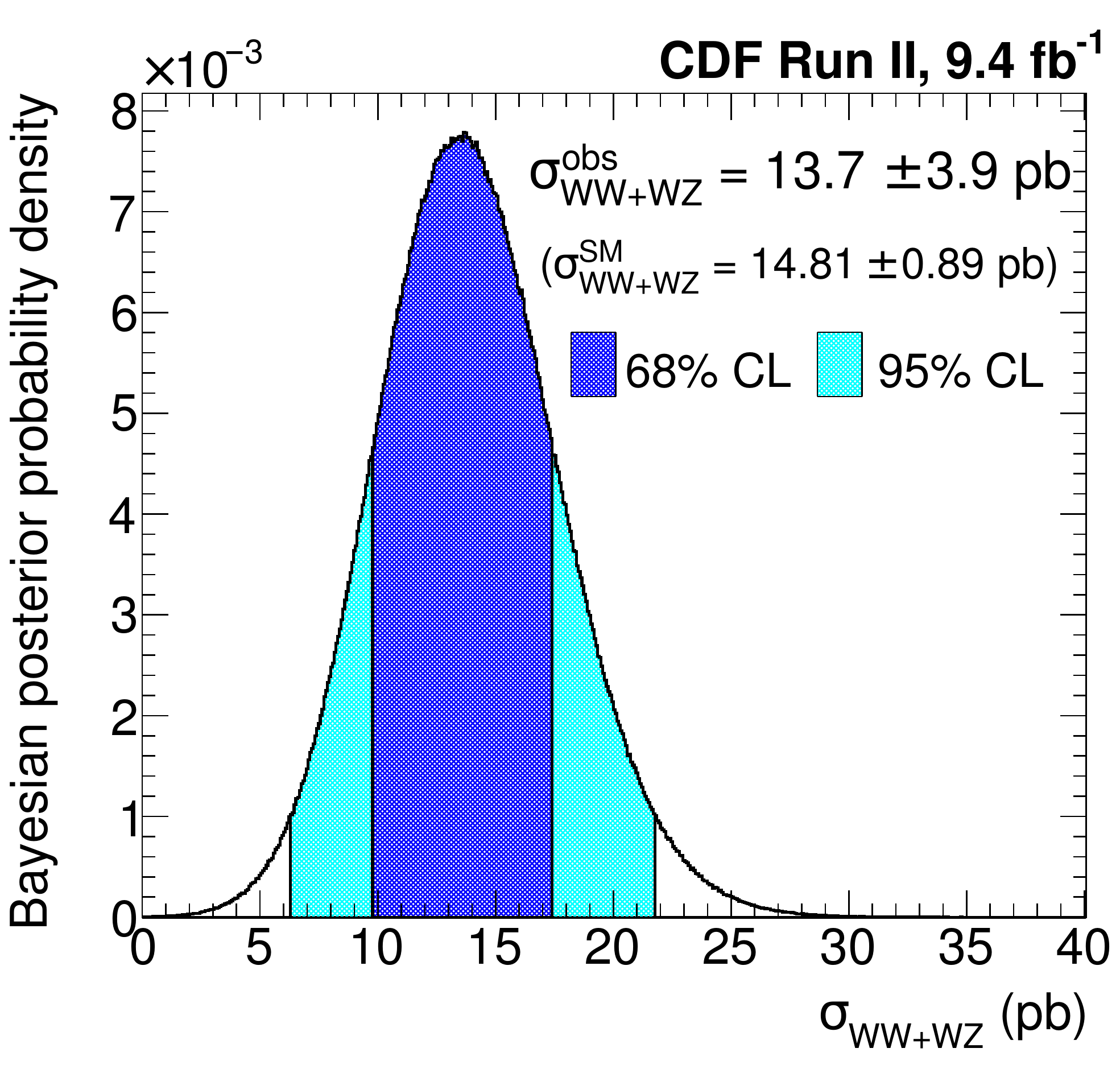}
\caption{Bayesian posterior distribution of the {\it WW+WZ} cross section after marginalization over the nuisance parameters. The maximum value is the measured cross-section value. The dark and light areas represent the smallest intervals enclosing  68.3\% and 95.5\% of the posterior integrals, respectively.}
\label{fig:posterior_dib}
\end{figure}

The resulting posterior distribution of the {\it WW+WZ} cross section is shown in Fig.~\ref{fig:posterior_dib} 
together with the 68.3\% and 95.5\% Bayesian credibility intervals. 
The measured signal strength of  $\mu^{\textrm{obs}}_{WW+WZ}=0.92\pm 0.26$  corresponds to a cross section of
\begin{equation}
  \sigma_{WW+WZ}^{\textrm{obs}} = 13.7 \pm 2.4 (\textrm{stat}) \pm 2.9 (\textrm{syst}) = 13.7\pm 3.9\textrm{~pb,} 
\end{equation}
in agreement with the SM NLO prediction of $\sigma_{WW+WZ}^{\textrm{SM}} = 14.8\pm 0.9$~pb~\cite{MCFM}.

\begin{figure} [!ht]
\centering
\includegraphics[width=0.49\textwidth]{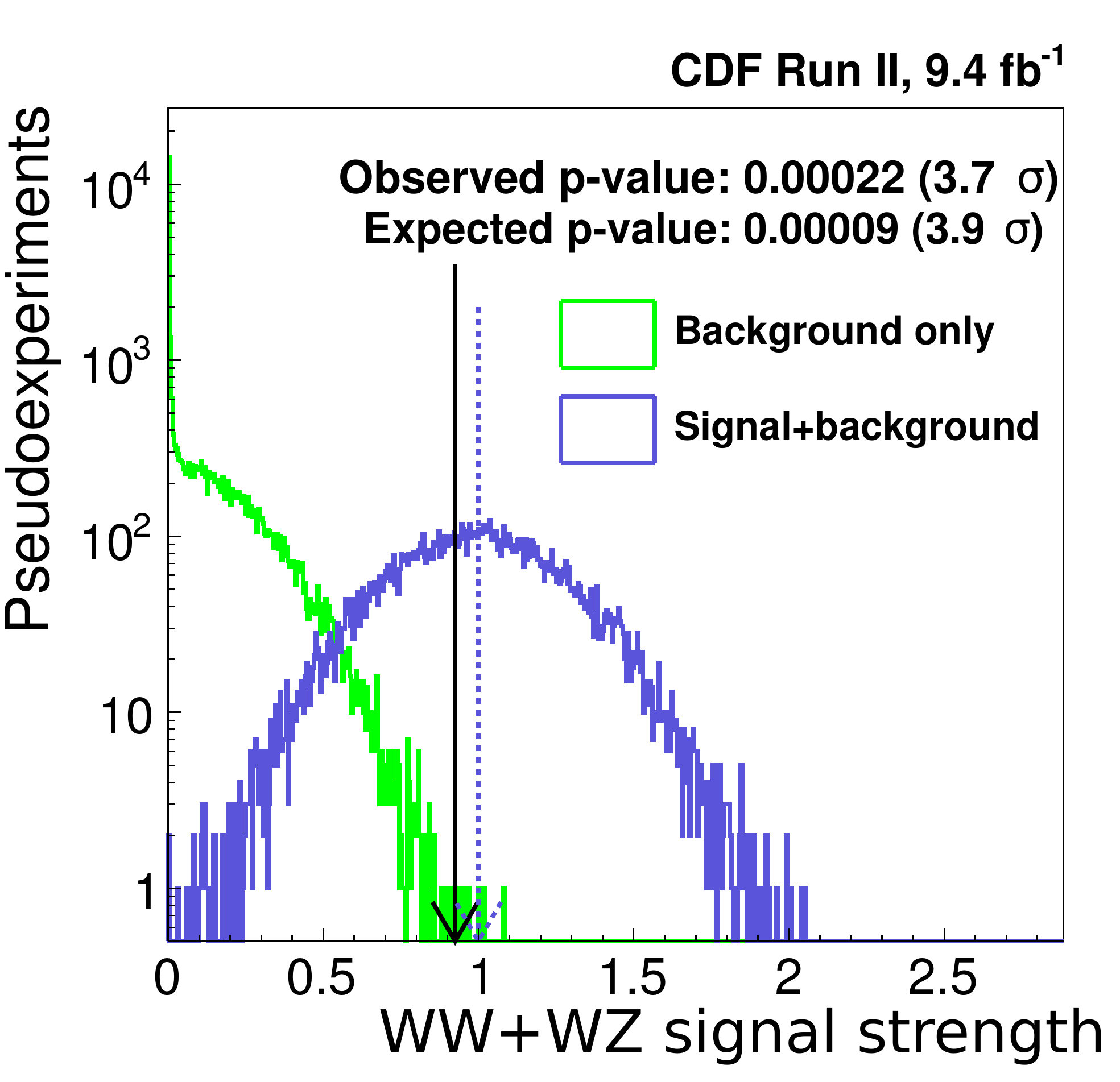}
\caption{Distribution of the results obtained in simplified simulated experiments  generated according to the background-only or the background-plus-signal hypothesis. 
The observed (expected) signal strength of $\mu^{\textrm{obs}}_{WW+WZ}=0.92$ ($\mu^{\textrm{exp}}_{WW+WZ}=1.0$) is indicated by the solid (dashed) arrow.}
\label{fig:significance_dib}
\end{figure}

To determine the significance of the signal, we perform a hypothesis test by comparing the data with
expectations under the null hypothesis of contributions from backgrounds only.  
Figure~\ref{fig:significance_dib} shows the distribution of the results obtained in simplified
simulated experiments that include contributions from signal and
background, and background only processes.
The probability $p_0$ of a background fluctuation to produce a signal strength equal or greater 
than the observed signal strength is $2.2\times 10^{-4}$,
corresponding to evidence for {\it WW}+{\it WZ} production in the $\ell\nu~$HF final-state with a 
significance of $3.7\sigma$. The result is compatible with the expected significance of $3.9\sigma$, 
obtained from pseudoexperiments generated under the SM hypothesis.

\begin{figure} [!ht]
\centering
\includegraphics[width=0.49\textwidth]{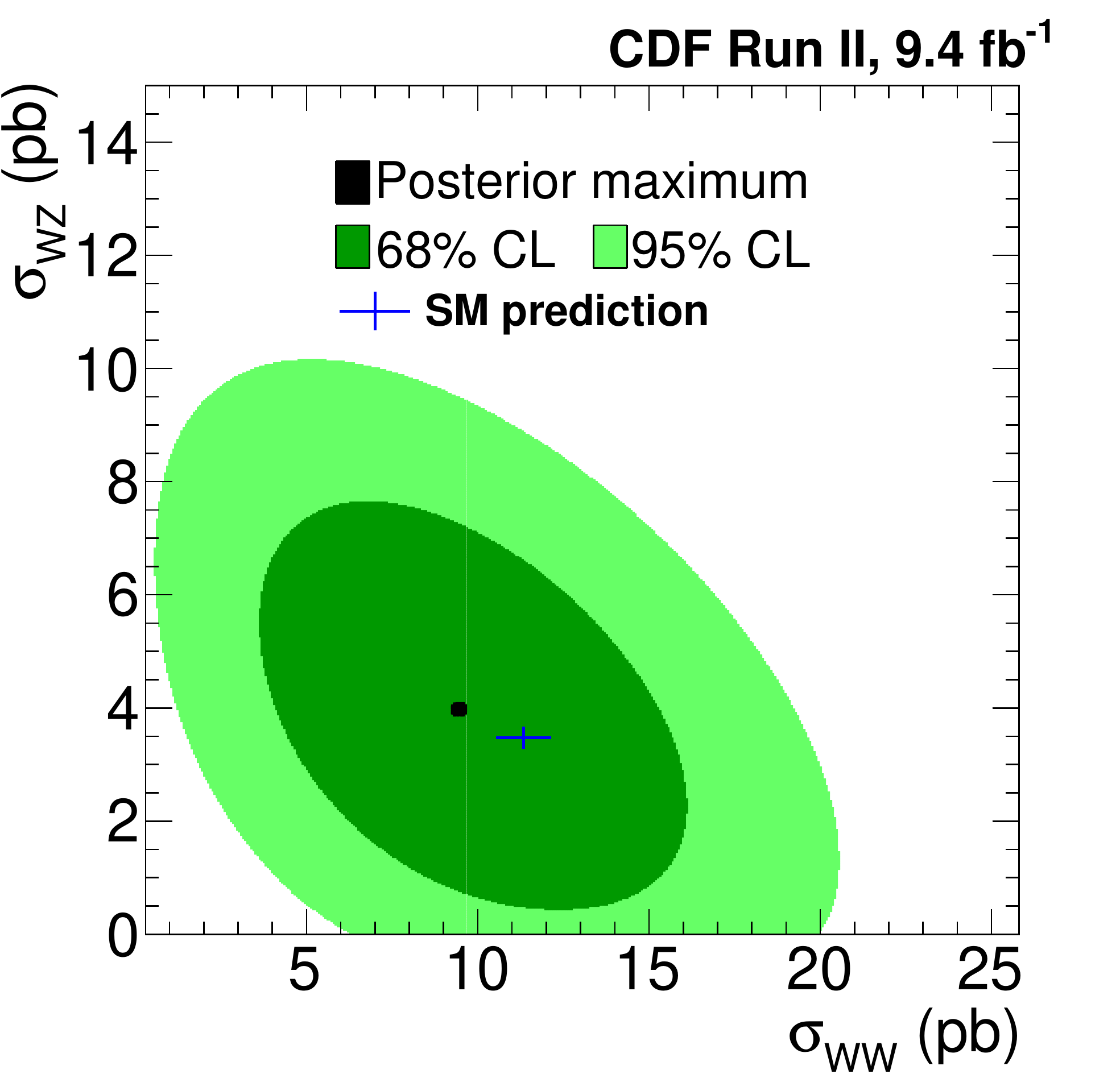}
\caption{Two-dimensional posterior probability distribution in the  $\sigma_{\it WW}$ and 
$\sigma_{\it WZ}$ plane, marginalized over the nuisance parameters. The measured values correspond 
to the maximum value of $\sigma^{\textrm{obs}}_{\it WW}=9.4$~pb and 
$\sigma^{\textrm{obs}}_{\it WZ}=3.7$~pb. 
The dark and light areas represent the smallest intervals enclosing 68.3\% and 95.5\% of the 
posterior integrals, respectively. The cross shows the SM predictions and the corresponding uncertainties~\cite{MCFM}.}
\label{fig:posterior_2d}
\end{figure}

The {\it WW} and {\it WZ} cross sections are also measured separately by exploiting the differing decay patterns of the $W$ and $Z$ bosons, which result in differing 
signal fractions in the one- and two-tag signal regions and in 
different distributions in the flavour separator NN.
The Bayesian analysis is repeated leaving the $\sigma_{\it WW}$ and $\sigma_{\it WZ}$ parameters 
free to vary.
Figure~\ref{fig:posterior_2d} shows the resulting Bayesian posterior distribution, 
with integration contours at 68.3\% and 95.5\%  Bayesian credibility levels. 
The maximum value corresponds to measured  signal strengths of $\mu^{\textrm{obs}}_{WW}=0.83$ and $\mu^{\textrm{obs}}_{WZ}=1.07$\,.
If the two-dimensional-Bayesian posterior is integrated with respect to the $\sigma_{WZ}$  or to the $\sigma_{WW}$ variable, 
the results are
\begin{eqnarray}
  \label{eq:ww_vs_wz}
  \sigma^{\textrm{obs}}_{WW} = &9.4 ^{+3.0}_{-3.0} (\textrm{stat}) ^{+2.9}_{-2.9} (\textrm{syst})& = 9.4\pm 4.2 \textrm{~pb,} \\
  \sigma^{\textrm{obs}}_{WZ} = &3.7^{+2.0}_{-1.8} (\textrm{stat}) ^{+1.4}_{-1.2} (\textrm{syst})& = 3.7^{+2.5}_{-2.2}\textrm{~pb,}  
\end{eqnarray}
in agreement with the SM NLO predictions of $\sigma_{WW}^{\textrm{SM}} = 11.3\pm 0.7$~pb and  
$\sigma_{WW}^{\textrm{SM}} = 3.5\pm 0.2$~pb~\cite{MCFM}, and corresponding to  the most precise 
measurement of the {\it WZ}-production 
cross section in a semileptonic final state to date. 
Although the expected value of $\sigma_{WW}^{\textrm{SM}}$ is about three times larger than the $\sigma_{WZ}^{\textrm{SM}}$ value, 
the relative uncertainties in the {\it WW} and {\it WZ} cross-section measurements are 
comparable because of the low $c$-jet identification efficiency and the larger systematic uncertainty associated with it. 
In addition, the {\it WZ}-signal yield is limited by the 15\% decay rate of the $Z$ boson to 
$b$-quark pairs.

\begin{figure*}[!ht]
\centering
\subfloat[]{\includegraphics[width=0.49\textwidth]{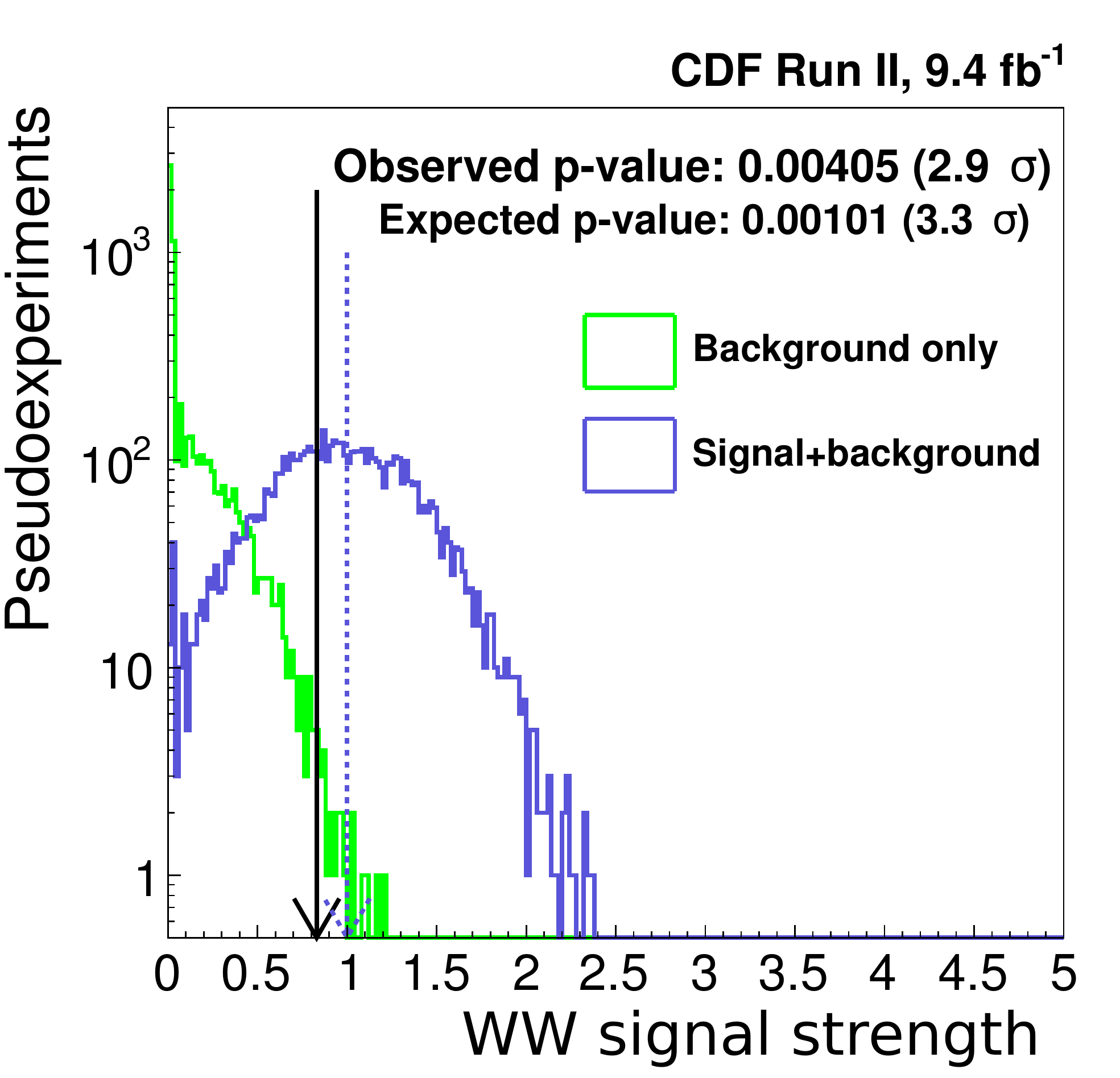}}
\subfloat[]{\includegraphics[width=0.49\textwidth]{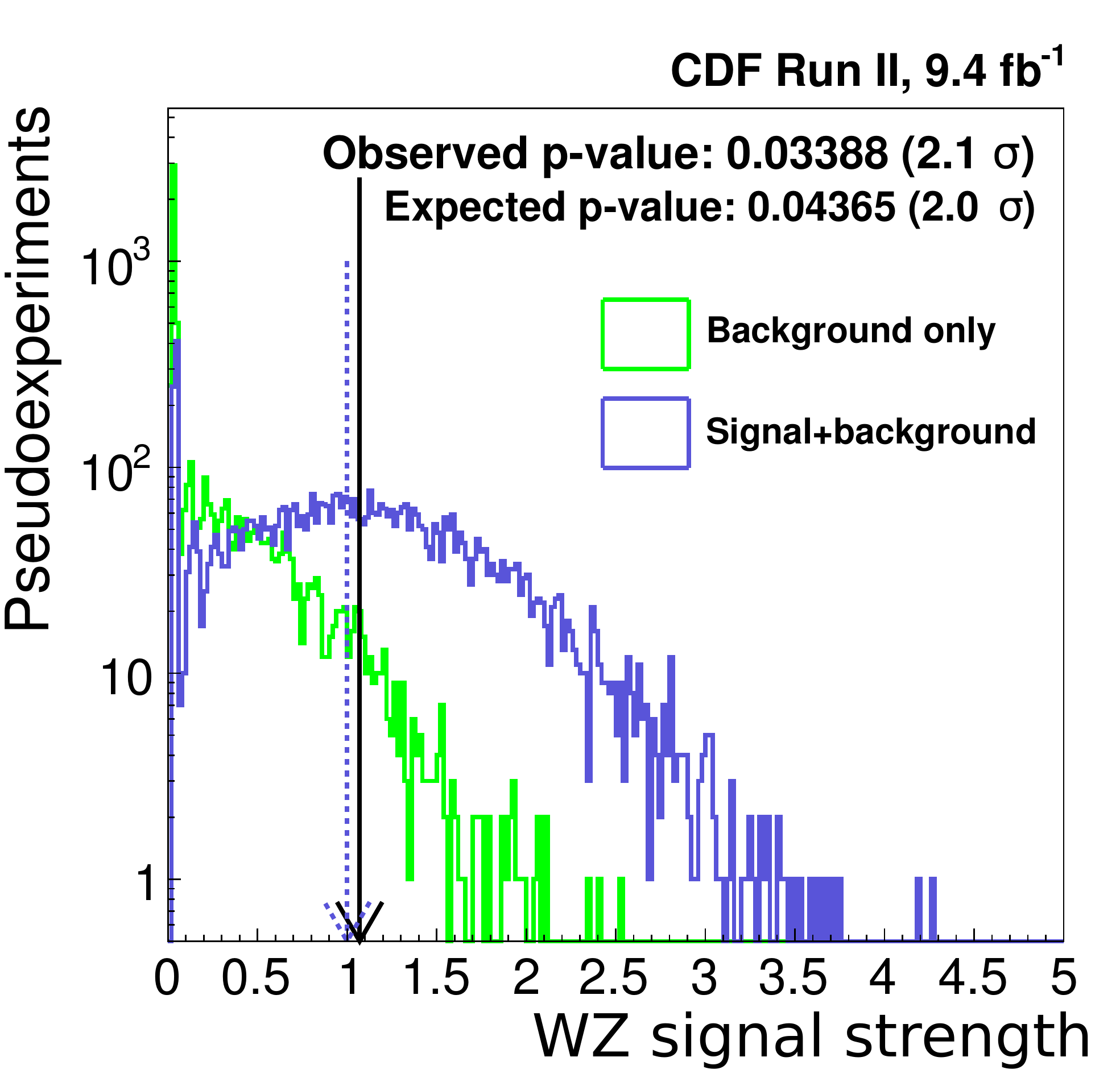}}
\caption{Distribution of  signal
strengths observed in simplified simulated experiments generated in a background-only or in a 
background-plus-signal 
hypothesis in the $\sigma_{WW}$ vs $\sigma_{WZ}$ plane and then integrated with respect to (a) $\sigma_{WZ}$ and  (b) $\sigma_{WW}$. 
The observed (expected) signal strengths are pointed by the solid (dashed) arrows.}
\label{fig:significance_2d_proj}
\end{figure*}

The separate significances of the {\it WW} and {\it WZ} signals are evaluated as done for the combined signal. 
Simulated experiments are generated under  the null hypothesis for both {\it WW} and {\it WZ} 
signals; then the cross sections 
measured on the $\sigma_{WW}$ vs $\sigma_{WZ}$ plane are integrated with respect to one or to the other variable and 
compared with $\sigma^{\textrm{obs}}_{WW}$ and $\sigma^{\textrm{obs}}_{WZ}$. 
The results of the $p_0$ estimates are reported in Fig.~\ref{fig:significance_2d_proj}. We obtain: 
$p_0^{WW}= 4.0\times 10^{-3}$ and $p_0^{WZ}=3.4 \times 10^{-2}$. These correspond to significances of $2.9\sigma$ and $2.1\sigma$, 
for {\it WW} and {\it WZ}, respectively. The result is consistent with the expected significances of $3.3\sigma$ and $2.0\sigma$, for {\it WW} and {\it WZ}, respectively.

The sensitivity of this analysis to the {\it WZ} process is similar to that
of the CDF {\it WH}  analysis in events with one electron or muon~\cite{WHbb_2012}, 
although the two analyses use different HF-tagging algorithms and different signal-to-background discrimination strategies. 

\section{Conclusions}\label{sec:conclusions}

We analyze the full proton-antiproton collision data set collected by the CDF experiment in Run II, 
corresponding to $9.4$~fb$^{-1}$\, of integrated luminosity, 
searching for the associated production of a $W$ boson decaying leptonically, and a $W$ or $Z$ 
boson decaying into heavy-flavored hadrons
($W\to cs$ and $Z\to c\bar{c},b\bar{b}$). 

Because of the small expected signal yield, the acceptance for $W\to\ell\nu$ events is 
maximized by using loose lepton-identification criteria, 
while reducing the multijet background using a multivariate SVM-based selection.

The signal  is identified by requiring two jets in each event, one or both consistent with being produced by heavy flavors, based on the presence of a secondary vertex.

The analysis of the dijet mass spectrum in the single- and double-tagged events shows 3.7$\sigma$ 
evidence for the {\it WW+WZ} signal over a background that is approximately 30 times larger.
The measured  total production cross-section of \mbox{$\sigma_{WW+WZ}=13.7 \pm 3.9$} pb is consistent with the 
SM predictions.

The {\it WW} and {\it WZ} processes are also investigated separately. 
An artificial neural network dedicated to distinguishing the
various flavors is used in conjunction with 
 the dijet mass spectrum to identify the different heavy-flavor-decay pattern of the $W$ and $Z$ bosons.
This gives an observed (expected) signal significance of $2.9\sigma$ (3.3$\sigma$) and $2.1\sigma$ (2.0$\sigma$) for {\it WW} and {\it WZ} processes, respectively. 
The measured total cross sections, within the full solid-angle acceptance, of $\sigma_{WW}=9.4\pm{4.2}$ pb and 
$\sigma_{WZ}=3.7^{+2.5}_{-2.2}$~pb are consistent with the SM expectations. 
The {\it WZ} cross section measurement is the most precise obtained to date with a semileptonic final state and 
 supports the CDF capability to identify rare
processes in this topology.

\section*{Acknowledgement}
We thank the Fermilab staff and the technical staffs of the
participating institutions for their vital contributions. This work
was supported by the U.S. Department of Energy and National Science
Foundation; the Italian Istituto Nazionale di Fisica Nucleare; the
Ministry of Education, Culture, Sports, Science and Technology of
Japan; the Natural Sciences and Engineering Research Council of
Canada; the National Science Council of the Republic of China; the
Swiss National Science Foundation; the A.P. Sloan Foundation; the
Bundesministerium f\"ur Bildung und Forschung, Germany; the Korean
World Class University Program, the National Research Foundation of
Korea; the Science and Technology Facilities Council and the Royal
Society, UK; the Russian Foundation for Basic Research; the Ministerio
de Ciencia e Innovaci\'{o}n, and Programa Consolider-Ingenio 2010,
Spain; the Slovak R\&D Agency; the Academy of Finland; and the
Australian Research Council (ARC).

\begin{appendix}
{ \renewcommand\thesection{}
\section{Simultaneous extraction of $W+c$, $W+c\bar{c}$, and $W+b\bar{b}$ normalization correction factors}\label{app:k}

As described in Section~\ref{sec:HF_bk}, a normalization correction factor, $K$, is used to account for 
the $W+$HF yield difference in data with respect to the HF fractions predicted by 
the \textsc{alpgen}~\cite{alpgen} simulation. 
Following Ref.~\cite{single_top}, the factor $K$ is extracted from the $W+1$ jet sample; however, in this analysis the 
$W+b\bar{b}$ plus $W+c\bar{c}$, and $W+c$ corrections are extracted simultaneously.

The $W+1$ jet selection is performed using a simpler selection compared to the main analysis:
$W\to \ell \nu $ candidates are selected using only the central-electron and central-muon trigger categories 
(described in Sec.~\ref{sec:OnlineSelection}) in combination with only the tight central-lepton 
identification algorithms (described in Sec.~\ref{sec:ev_sel}). 
The one-jet-pretag selection region is then defined by the presence of exactly one jet 
of \Et$>$20 GeV and $|\eta|<2.0$ in each event, and multijet background contamination is reduced by means 
of a selection requirement on the central-region SVM-output-value distribution, 
as described in Sec.~\ref{sec:multijet_sel}.
The sample is then enriched in HF jets by requiring the jet to be tagged by the tight \textsc{secvtx} algorithm. 
The background estimate proceeds as described in Sec.~\ref{sec:Back} for the $W+2$ jets sample, 
with the notable difference that the signals under investigation are now $W+b\bar{b}$ plus $W+c\bar{c}$, 
and $W+c$ production.

\begin{figure}[!h]
  \centering
  \includegraphics[width=0.49\textwidth]{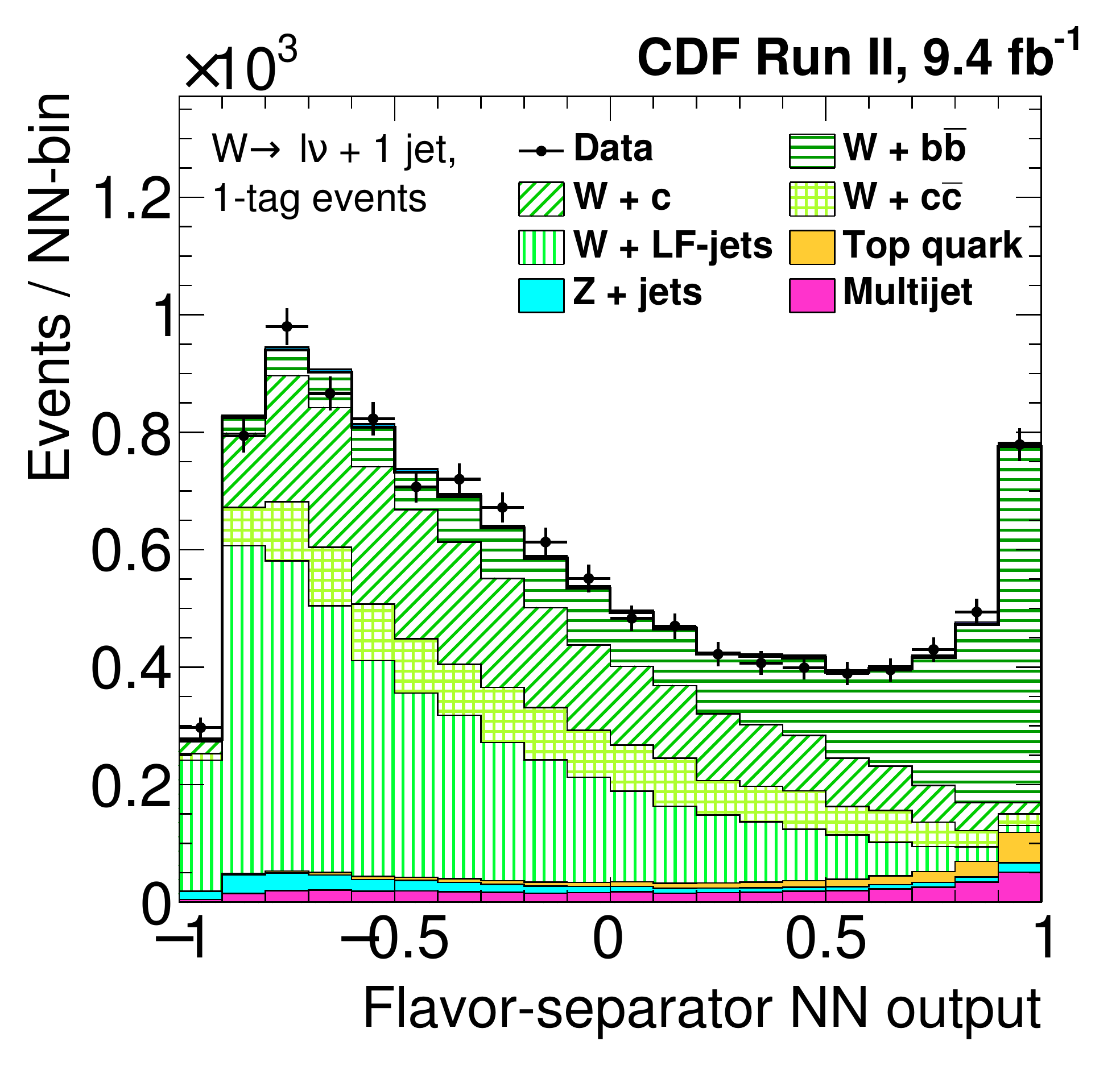}
  \caption{Flavor-separator-NN output distribution for the leptonically decaying $W$-plus-one-tagged-jet events; 
 central-muon and central-electron categories are combined.} 
 \label{fig:KIT_postfit}
\end{figure}

The signal strengths of the  $W+b\bar{b}$ plus $W+c\bar{c}$, and $W+c$  processes are defined as the ratio between 
the input HF-correction factors, $K_{bb,cc}=1.0$ and $K_c = 1.0$, and the ones favored by the data. 
We compare the flavor-separator-NN output distribution,  shown in Fig.~\ref{fig:KIT_postfit}, of the two signals 
and of the backgrounds to the one obtained from data using the same Bayesian methodology described Sec.~\ref{sec:fit}. 
A two-dimentional Bayesian posterior distribution is used to extract simultaneously the two signal strengths.

\begin{figure}[!h]
  \centering
  \includegraphics[width=0.49\textwidth]{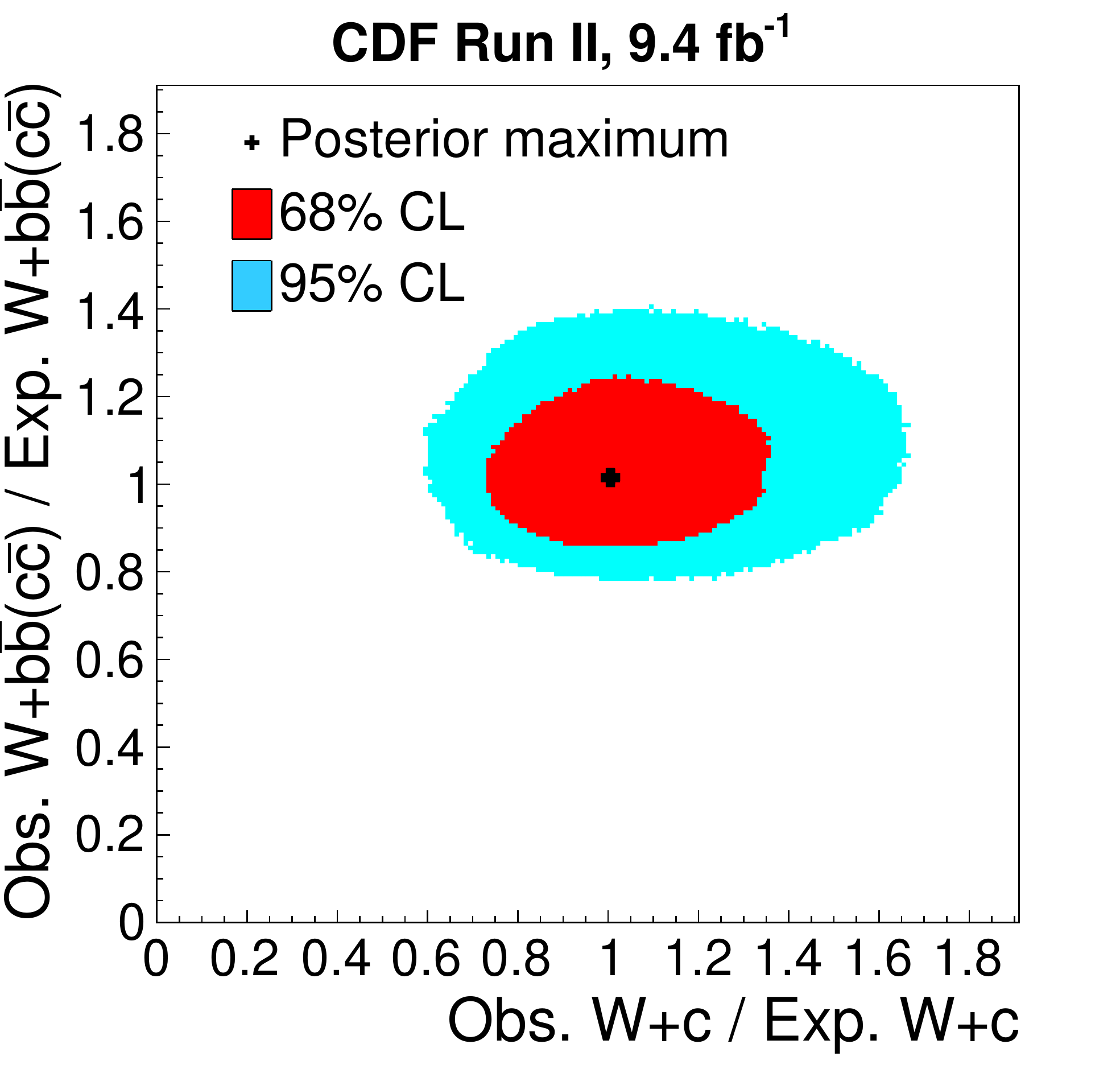}
  \caption{Final two-dimensional  posterior distribution of $W+b\bar{b}$ plus $W+c\bar{c}$ versus $W+c$ signal strengths. 
The signal strengths come from the ratio between the input HF correction factors used during the background estimate 
and the one favored by the data.} \label{fig:Kfact_posterior}
\end{figure}

The analysis procedure is iterated four times, each time scaling the HF correction factor by the signal strength 
extracted from the  previous iteration. The signal strength given by the last iteration is consistent with one, 
as shown by the two-dimensional-Bayesian posterior distribution in Fig.~\ref{fig:Kfact_posterior},
and the final HF correction factors are  \mbox{$K_{bb,cc} = 1.24\pm 0.25$}, and $K_c = 1.0\pm 0.3$. 

}
\end{appendix}
%\clearpage
\bibliographystyle{plain}
%\bibliography{references}

\begin{thebibliography}{100}
\bibitem{d0} The track impact parameter $d_0$ is defined as the distance 
from the beam-line axis to the track at the track's closest approach in the transverse plane.
\bibitem{UE_explain} The underlying event consists of the beam-beam 
remnants and the multiple-parton interactions that accompany a
hard scattering.
\end{thebibliography}

\end{document}